\documentclass{jpp}

\usepackage{graphicx}
\usepackage{natbib}

\usepackage{graphics}
\usepackage{epsfig}
\usepackage{bm}
\usepackage{amsmath,amssymb}
\usepackage{stmaryrd}

\def\ov#1{\overline{#1}}

\def\vb#1{\mbox{\boldmath$#1$}}
\def\pd#1#2{\frac{\partial #1}{\partial #2}}
\def\fd#1#2{\frac{\delta #1}{\delta #2}}
\def\wh#1{\widehat{#1}}
\def\bdot{\,\vb{\cdot}\,}
\def\btimes{\,\vb{\times}\,}

\def\bhat{\wh{{\sf b}}}
\def\cal#1{\mathcal{#1}}

\def\bhat{\wh{{\sf b}}}

\newcommand{\bc}{\begin{center}}
\newcommand{\ec}{\end{center}}
\newcommand{\bt}{\begin{tabbing}}
\newcommand{\et}{\end{tabbing}}
\newcommand{\be}{\begin{equation}}
\newcommand{\ee}{\end{equation}}
\newcommand{\ba}{\begin{eqnarray}}
\newcommand{\ea}{\end{eqnarray}}

\title[Exact conservation laws for gauge-free electromagnetic gyrokinetic equations]{Exact conservation laws for gauge-free electromagnetic gyrokinetic equations}

\author[Brizard]%
{Alain J.~Brizard%
  \thanks{Email address for correspondence: abrizard@smcvt.edu}}

\affiliation{Department of Physics, Saint Michael's College, Colchester, VT 05439, USA}

\date{\today; revised ?; accepted ?. - To be entered by editorial office}
\begin{document}

\maketitle

\begin{abstract} 
The exact energy and angular-momentum conservation laws are derived by Noether method for the Hamiltonian and symplectic representations of the gauge-free electromagnetic gyrokinetic Vlasov-Maxwell equations. These gyrokinetic equations, which are solely expressed in terms of electromagnetic fields, describe the low-frequency turbulent fluctuations  that perturb a time-independent toroidally-axisymmetric magnetized plasma. The explicit proofs presented here provide a complete picture of the transfer of energy and angular momentum between the gyrocenters and the perturbed electromagnetic fields, in which the crucial roles played by gyrocenter polarization and magnetization effects are highlighted. In addition to yielding an exact angular-momentum conservation law, the gyrokinetic Noether equation yields an exact momentum transport equation, which might be useful in more general equilibrium magnetic geometries.
\end{abstract}

\section{Introduction}

Nonlinear gyrokinetic theory has been at the forefront of plasma physics research since the pioneering work of \cite{Frieman_Chen_1982}. In its modern representation \citep{Brizard_Hahm_2007}, nonlinear gyrokinetic theory involves a two-step (guiding-center + gyrocenter) transformation leading to the adiabatic invariance of the gyrocenter magnetic moment and a set of reduced gyrocenter Hamilton equations that is decoupled from the fast gyromotion dynamics.  The guiding-center and gyrocenter dynamical reductions, on the other hand, introduce guiding-center and gyrocenter polarization and magnetization in the gyrokinetic Maxwell equations, which play crucial roles in the self-consistent evolution of a turbulent magnetized plasma.

The gyrokinetic Vlasov-Maxwell equations presented in this work are based on the gauge-free electromagnetic-field gyrokinetic formulation recently introduced by \cite{Burby_Brizard_2019} and \cite{Brizard_2020}, in which only the perturbed electromagnetic fields appear in the gyrokinetic Vlasov-Maxwell equations. This gyrokinetic electromagnetic-field formulation, which also facilitates the development of hybrid kinetic particle simulation schemes \citep{Chen_Parker_2009,Chen_2019}, has been a topic of recent research interest \citep{Chen_2020}.

\subsection{Gyrokinetic energy conservation law}

The energy conservation laws of several sets of gyrokinetic equations have been the topic of active research since the Hamiltonian gyrokinetic work of \cite{Dubin_1983}, where the energy conservation law was derived directly from the electrostatic gyrokinetic Vlasov-Poisson equations. Next, the energy conservation laws for the finite-beta electromagnetic gyrokinetic equations and the fully electromagnetic gyrokinetic equations were constructed by \cite{HLB_1988} and \cite{Brizard_1989a}, respectively, and then proved explicitly by \cite{Brizard_1989b} for the fully electromagnetic case.  Since its inception, the primary use of an exact energy conservation law associated with a gyrokinetic model has been as a measure of the accuracy of its numerical implementation using gyrokinetic particle simulation methods \citep{Garbet_2010}.

The discovery of the Lagrangian \citep{Sugama_2000}, Eulerian \citep{Brizard_PRL_2000,Brizard_PoP_2000}, and Euler-Poincar\'{e} \citep{Squire_2013} variational formulations for the nonlinear gyrokinetic equations led to the direct derivations of exact gyrokinetic energy conservation laws by Noether method \citep{Goldstein_2002,Brizard_2005b}. While the Noether derivation guarantees the existence of an exact energy conservation law, its explicit proof often provides useful insights into the transfer of energy between the perturbed electromagnetic fields, on the one hand, and the gyrocenter Vlasov distribution, on the other hand. This is especially important since polarization and magnetization combines particles and fields at all orders in the gyrocenter perturbation analysis. Using an Eulerian variational formulation, explicit proofs were also presented for several nonlinear gyrokinetic models \citep{Brizard_1999,Brizard_2010a} and reduced-fluid plasma models \citep{SSB_2005,Brizard_2005a,Brizard_NFLR_2008}.

\subsection{Gyrokinetic angular-momentum conservation law}

The topic of an exact gyrokinetic momentum conservation law gained crucial importance in the context of the momentum transport \citep{Waltz_2007,Parra_2010,Abiteboul_2011,Peeters_2011} and intrinsic toroidal rotation \citep{Wang_Peng_2018,Stoltzfus-Dueck_2019} in axisymmetric tokamak plasmas. In particular, the phenomenon of intrinsic toroidal rotation, which is observed in the absence of external torque, must be investigated within the context of toroidal angular-momentum conservation. \cite{Scott_Smirnov_2010} derived an explicit toroidal angular-momentum conservation law for the electrostatic gyrokinetic equations by deriving it as a moment of the gyrokinetic Vlasov equation. The same equation was rederived by Noether method, and explicitly shown to be exact, by \cite{Brizard_Tronko_2011}, while the Noether derivation of the momentum conservation law was also considered for several reduced plasma fluid models \citep{Brizard_2005b,Brizard_2010b}. 

\subsection{Previous variational derivations of gyrokinetic conservation laws}
 
 The angular-momentum conservation law in gyrokinetic Vlasov-Maxwell models have regained significant interest recently in several works by \cite{EH_2020}, \cite{Fan_2020}, and \cite{Sugama_2021}. By not splitting the magnetic field into equilibrium (time-independent) and perturbed (time-dependent) components, both \cite{Fan_2020} and \cite{Sugama_2021} rederived the guiding-center energy-momentum conservation laws derived earlier by \cite{Sugama_2016}, using a direct moment approach of the drift-kinetic equation, and by \cite{Brizard_Tronci_2016}, using several equivalent guiding-center variational principles. \cite{Fan_2020} generalized earlier results by \cite{Pfirsch_Morrison_1985} and \cite{Similon_1985} by including higher-order guiding-center gyrogauge corrections.

In particular, using a hybrid gyrokinetic model that includes gyrokinetic electrostatic fluctuations with a weakly time-dependent guiding-center Lagrangian, \cite{Sugama_2021} obtained a symmetric gyrokinetic stress tensor $T^{ji} = T^{ij}$ by using a formula $T^{ij} \equiv {\cal L}\,g^{ij} - 2\,\partial{\cal L}/\partial g_{ij}$ obtained from the general theory of relativity \citep{Landau_1971}, in which partial derivatives of the Lagrangian density ${\cal L}$ with respect to components of the symmetric metric tensor ${\sf g}$ are evaluated (here, a specific choice for the spatial coordinates is not needed). Additional comments about the works of \cite{Fan_2020} and \cite{Sugama_2021} will be made below in Sec.~\ref{sec:symp_gyro}.

The work of \cite{EH_2020} presents an Euler-Poincar\'{e} variational principle for the drift-kinetic limit of the gauge-free gyrokinetic Vlasov-Maxwell model of \cite{Burby_Brizard_2019}. In this work, the standard gyrokinetic separation of equilibrium and perturbed components for the electromagnetic field is used, and the asymmetry of the resulting gyrokinetic stress tensor is shown to be driven by electromagnetic-field perturbations only. The exact conservation law of toroidal angular momentum, however, will be guaranteed under the assumption of an axisymmetric equilibrium magnetic field. The energy-momentum conservation laws derived by \cite{EH_2020}, which will be rederived here from an Eulerian variational principle, will be explicitly proved and expanded in the present work.

\subsection{Organization}

The remainder of the present paper is organized as follows. In Sec.~\ref{sec:gc_VM}, we review the work of \cite{Brizard_2008} where the exact energy-momentum conservation laws are given for a generic set of reduced Vlasov-Maxwell equations, in which the electromagnetic fields are not separated into background and perturbed parts. The paradigm set of reduced plasma equations is given by the guiding-center Vlasov-Maxwell equations, with variational formulations \citep{Pfirsch_Morrison_1985,Similon_1985,Brizard_Tronci_2016} leading to exact reduced energy-momentum conservation laws. While the resulting reduced stress tensor is manifestly asymmetric, as noted by \cite{Pfirsch_Morrison_1985} and \cite{Similon_1985}, we show that the apparent asymmetry of the reduced stress tensor is due to polarization and magnetization effects derived from a ponderomotive Hamiltonian \citep{Cary_Kaufman_1981,Brizard_2009}. Since the stress tensor must be explicitly symmetric when the electromagnetic fields $({\bf E},{\bf B})$ are not split into time-independent (equilibrium) and time-dependent (perturbed) components,  we use the guiding-center Vlasov-Maxwell model of \cite{Brizard_Tronci_2016} and show that the explicit expressions of the guiding-center polarization and magnetization guarantee a symmetric guiding-center stress tensor. 

In Sec.~\ref{sec:symp_gyro}, we review the gauge-free gyrocenter Hamiltonian models derived by \cite{Burby_Brizard_2019} and \cite{Brizard_2020}, where the equations of motion are solely expressed in terms of the perturbed electromagnetic fields 
$({\bf E}_{1},{\bf B}_{1})$. Here, following the standard gyrokinetic formalism \citep{Brizard_Hahm_2007}, the magnetic field ${\bf B} = {\bf B}_{0} + \epsilon\,{\bf B}_{1}$ is split into the time-independent equilibrium magnetic field ${\bf B}_{0}$, which is assumed to be axisymmetric (i.e., $\partial{\bf B}_{0}/\partial\varphi = \wh{\sf z}\btimes{\bf B}_{0}$), and the time-dependent magnetic-field perturbation ${\bf B}_{1}$ ($\epsilon$ denotes the magnitude of the perturbation). In addition, we assume that ${\bf E} = \epsilon\,{\bf E}_{1}$ appears solely as a perturbation electric field in the present work, although a equilibrium electric field may also be considered \citep{Brizard_1995}. We note that the choice of the magnetic perturbation ${\bf B}_{1}$ is consistent with the source-free perturbed Maxwell equations $\nabla\bdot{\bf B}_{1} = 0$ and $\partial{\bf B}_{1}/\partial t = -\,c\,\nabla\btimes{\bf E}_{1}$.  In Sec.~\ref{sec:symp_VP}, the gyrokinetic Vlasov-Maxwell equations are derived from a variational principle \citep{Brizard_PRL_2000,Brizard_PoP_2000}, from which explicit expressions for the gyrocenter polarization and magnetization are obtained for both gauge-free gyrokinetic models. In Sec.~\ref{sec:symp_laws}, the gyrokinetic conservation laws are derived by Noether method. In the present work, we will show that the apparent asymmetry of the gyrokinetic stress tensor, which is only due to electromagnetic-field perturbations (since the guiding-center stress tensor is symmetric), plays a crucial role in establishing an exact toroidal angular-momentum conservation law in the presence of a non-uniform (but axisymmetric) equilibrium magnetic field. We will also show that an exact gyrokinetic momentum transport equation will be obtained from the gyrokinetic Noether equation, which can be used to study momentum transport in general magnetic geometries.

\section{\label{sec:gc_VM}Conservation laws for the reduced Vlasov-Maxwell equations}

Although our primary motivation is to discuss exact conservation laws of gyrokinetic systems, we present a brief discussion on the form of these conservation laws for a generic set of reduced Vlasov-Maxwell equations, based on work presented at the 2006 Vlasovia conference \citep{Brizard_2008}. In this formal derivation, the electromagnetic fields $({\bf E} = -\nabla\Phi - c^{-1}\partial_{t}{\bf A},\,{\bf B} = \nabla\btimes{\bf A})$ are not split into time-independent background and time-dependent perturbed components and, by using canonical coordinates $(\ov{\bf x}, \ov{\bf p})$, the dependence on the potentials $(\Phi,
{\bf A})$ and the fields $({\bf E},{\bf B})$ only enter through the reduced Hamiltonian 
\begin{equation} 
\ov{H}(\ov{\bf p};\Phi,{\bf A},{\bf E},{\bf B}) \;\equiv\; m\,|\ov{\bf v}|^{2}/2 \;+\; e\,\Phi \;+\; \ov{\Psi}(\ov{\bf v}; {\bf E}, {\bf B}),
\label{eq:red_Ham}
\end{equation}
where $\ov{\bf v} \equiv [\ov{\bf p} - (e/c){\bf A}]/m$, the potentials and fields $(\Phi,{\bf A},{\bf E},{\bf B})$ are evaluated at the reduced position $\ov{\bf x}$, and $\ov{\Psi}(\ov{\bf v}; {\bf E}, {\bf B})$ denotes the {\it ponderomotive} Hamiltonian \citep{Cary_Kaufman_1981,Brizard_2009}. Here, the dependence of the ponderomotive Hamiltonian on the gauge-free term $\ov{\bf v}$ maintains the gauge-transformation property of the reduced Hamiltonian.

\subsection{Reduced Vlasov-Maxwell equations}

Using the reduced Hamiltonian \eqref{eq:red_Ham}, the reduced Vlasov-Maxwell equations are now expressed as follows. First, the reduced equations of motion are given in Hamiltonian canonical form as
\begin{eqnarray}
d\ov{\bf x}/dt = \partial\ov{H}/\partial\ov{\bf p} &=& \ov{\bf v} \;+\; \partial\ov{\Psi}/\partial\ov{\bf p}, \label{eq:red_x} \\
d\ov{\bf p}/dt = -\,\ov{\nabla}\ov{H} &=& -\,e\,\ov{\nabla}\Phi \;+\; (e/c)\,\ov{\nabla}{\bf A}\bdot\ov{\bf v} \;-\; \ov{\nabla}\ov{\Psi}. \label{eq:red_p}
\end{eqnarray}
If the reduced force equation \eqref{eq:red_p} is written in terms of $\ov{\bf v}$, we find
\begin{equation}
m\,\frac{d\ov{\bf v}}{dt} \;=\; e\,{\bf E} \;+\; \frac{e}{c}\frac{d\ov{\bf x}}{dt}\btimes{\bf B} \;+\; \ov{\nabla}{\bf E}\bdot\ov{\vb{\pi}} + \ov{\nabla}{\bf B}\bdot\ov{\vb{\mu}},
\label{eq:red_v}
\end{equation}
where we used Eq.~\eqref{eq:red_x} on the right side and the ponderomotive force
\[ -\,\ov{\nabla}\ov{\Psi} \;=\; -\,\ov{\nabla}{\bf A}\bdot\left(-\,\frac{e}{c}\pd{\ov{\Psi}}{\ov{\bf p}}\right) \;+\; \ov{\nabla}{\bf E}\bdot\ov{\vb{\pi}} \;+\; \ov{\nabla}{\bf B}\bdot\ov{\vb{\mu}} \]
includes the reduced electric and magnetic dipole moments $(\ov{\vb{\pi}}, \ov{\vb{\mu}}) \equiv (-\,\partial\ov{\Psi}/\partial{\bf E}, -\,\partial\ov{\Psi}/\partial{\bf B})$ derived from the ponderomotive Hamiltonian. The reduced Vlasov equation is, therefore,  expressed as
\begin{equation}
\pd{\ov{f}}{t} \;=\; -\,\frac{d\ov{\bf x}}{dt}\bdot\ov{\nabla}\ov{f} \;-\; \frac{d\ov{\bf p}}{dt}\bdot\pd{\ov{f}}{\ov{\bf p}}.
\label{eq:red_Veq}
\end{equation}
The reduced Maxwell equations, on the other hand, are expressed as
\begin{eqnarray}
\nabla\bdot{\bf E} & = & 4\pi\,\varrho \;\equiv\; 4\pi\,\left(\ov{\varrho} \;-\; \nabla\bdot\ov{\mathbb{P}}\right), \label{eq:red_divE} \\
\nabla\btimes{\bf B} - \frac{1}{c}\pd{\bf E}{t} & = & \frac{4\pi}{c}\,{\bf J} \;\equiv\; \frac{4\pi}{c} \left( \ov{\bf J} \;+\; \pd{\ov{\mathbb{P}}}{t} \;+\; c\,\nabla\btimes\ov{\mathbb{M}} \right), \label{eq:red_curlB}
\end{eqnarray}
with the source-free Maxwell equations
\begin{equation}
\left. \begin{array}{rcl}
\partial{\bf B}/\partial t \;+\; c\,\nabla\btimes{\bf E} & = & 0 \\
\nabla\bdot{\bf B} & = & 0
\end{array} \right\}.
\label{eq:free_Max}
\end{equation}
In Eqs.~\eqref{eq:red_divE}-\eqref{eq:red_curlB}, the reduced charge and current densities $(\ov{\varrho}, \ov{\bf J})$ and the reduced polarization and magnetization $(\ov{\mathbb{P}},\ov{\mathbb{M}})$ are derived from the reduced Hamiltonian:
\begin{equation}
\int_{\ov{\bf p}} \ov{f}\;\partial\ov{H}(\ov{\bf p};\Phi,{\bf A},{\bf E},{\bf B}) \;\equiv\; \partial\Phi\;\ov{\varrho} \;-\; \partial{\bf A}\bdot\ov{\bf J}/c \;-\; \partial{\bf E}\bdot\ov{\mathbb{P}} \;-\; \partial{\bf B}\bdot\ov{\mathbb{M}},
\label{eq:partial_H}
\end{equation}
where the notation $\int_{\ov{\bf p}}$ indicates an integral over canonical-momentum space (as well as including a sum over particle species), and $\partial$ denotes either a space-time partial derivative $(\nabla, \partial/\partial t)$ or an Eulerian variation $\delta$. Specifically, we find the definitions
\begin{equation}
\left(\ov{\rho}, \ov{\bf J}, \ov{\mathbb{P}} , \ov{\mathbb{M}} \right) \;\equiv\; \int_{\ov{\bf p}} \ov{f} \left( e,\; e\,\frac{d\ov{\bf x}}{dt},\; -\,\pd{\ov{\Psi}}{\bf E},\; -\,\pd{\ov{\Psi}}{\bf B}\right),
\end{equation}
where contributions arise from reduced particles located at the field position (i.e., $\ov{\bf x} = {\bf x})$.

We also note that the reduced Maxwell equations \eqref{eq:red_divE}-\eqref{eq:red_curlB} can be written in terms of the reduced Maxwell fields
\begin{equation}
\left. \begin{array}{rcl}
\ov{\mathbb{D}} & = & {\bf E} \;+\; 4\pi\,\ov{\mathbb{P}} \\
\ov{\mathbb{H}} & = & {\bf B} \;-\; 4\pi\,\ov{\mathbb{M}}
\end{array} \right\},
\label{eq:red_DH}
\end{equation}
as
\begin{eqnarray}
\nabla\bdot\ov{\mathbb{D}} & = & 4\pi\,\ov{\varrho}, \label{eq:red_divD} \\
\nabla\btimes\ov{\mathbb{H}} - \frac{1}{c}\pd{\ov{\mathbb{D}}}{t} & = & \frac{4\pi}{c}\, \ov{\bf J}, \label{eq:red_curlH}
\end{eqnarray}
which guarantees that the reduced charge conservation law 
\begin{equation}
\pd{\varrho}{t} + \nabla\bdot{\bf J} \;=\; \pd{}{t}\left(\ov{\varrho} \;-\; \nabla\bdot\ov{\mathbb{P}}\right) + \nabla\bdot\left( \ov{\bf J} + \pd{\ov{\mathbb{P}}}{t} + c\,\nabla\btimes\ov{\mathbb{M}} \right) \;=\; \pd{\ov{\varrho}}{t} \;+\; \nabla\bdot\ov{\bf J} \;=\; 0
\label{eq:red_charge}
\end{equation}
follows directly from the charge conservation law.

\subsection{Reduced energy-momentum conservation laws}

Since the electromagnetic field $({\bf E},{\bf B})$ is not split into equilibrium and perturbed components in the reduced Vlasov-Maxwell equations \eqref{eq:red_Veq}-\eqref{eq:red_curlB}, the energy-momentum conservation laws derived for the reduced Vlasov equation \eqref{eq:red_Veq} and the reduced Maxwell equations \eqref{eq:red_divE}-\eqref{eq:red_curlB} [or \eqref{eq:red_divD}-\eqref{eq:red_curlH}], with the source-free Maxwell equations \eqref{eq:free_Max}, are direct consequences of the Noether Theorem. Here, the reduced Noether equation \citep{Brizard_2008} is first expressed as
\begin{eqnarray}
0 &=& \pd{}{t} \left[ \int_{\ov{\bf p}} \ov{f}\;\delta\ov{\cal S} \;-\; \delta{\bf A}\bdot\frac{\ov{\mathbb{D}}}{4\pi c} \;+\;  \frac{\delta t}{8\pi}\;\left( |{\bf E}|^{2} \;-\frac{}{} |{\bf B}|^{2} \right) \right] \nonumber \\
 &&+\; \nabla\bdot\left[ \int_{\ov{\bf p}} \ov{f}\,\frac{d\ov{\bf x}}{dt}\;\delta\ov{\cal S} \;-\; \frac{1}{4\pi}\;\left(\delta\Phi\;\ov{\mathbb{D}} \;+\frac{}{} \delta{\bf A}\btimes\ov{\mathbb{H}}\right) \;+\; \frac{\delta{\bf x}}{8\pi}\;\left( |{\bf E}|^{2} \;-\frac{}{} |{\bf B}|^{2} \right) \right],
\label{eq:red_Noether}
 \end{eqnarray}
 where the Eulerian variations 
 \begin{equation}
 \left. \begin{array}{rcl}
 \delta\ov{\cal S} &=& \ov{\bf p}\bdot\delta{\bf x} - \ov{H}\,\delta t \\
 \delta\Phi &=& {\bf E}\bdot\delta{\bf x} - c^{-1}\partial\delta\chi/\partial t \\
 \delta{\bf A} &=& {\bf E}\;c\,\delta t + \delta{\bf x}\btimes{\bf B} + \nabla\delta\chi
 \end{array} \right\}, 
 \label{eq:Euler-SPhiA}
 \end{equation}
 with the gauge term defined as $\delta\chi \equiv \Phi\,c\,\delta t - {\bf A}\bdot\delta{\bf x}$, are generated by the space-time virtual displacements $(\delta{\bf x}, \delta t)$. We note that the gauge term $\delta\chi$ appears naturally when the Euler variations $\delta\Phi = -\,\delta t\,\partial\Phi/\partial t - \delta{\bf x}\bdot\nabla\Phi$ and $\delta{\bf A} = -\,\delta t\,\partial{\bf A}/\partial t - \delta{\bf x}\bdot\nabla{\bf A}$ are expressed in terms of the electric field ${\bf E} = -\nabla\Phi - c^{-1}\partial{\bf A}/\partial t$ and the magnetic field ${\bf B} = \nabla\btimes{\bf A}$.
 
 We now remove the gauge-dependent terms by using the identity
 \[ -\pd{}{t}\left(\nabla\delta\chi\vb{\cdot}\ov{\mathbb{D}}\right) + \nabla\vb{\cdot}\left(\pd{\delta\chi}{t}\;\ov{\mathbb{D}} - c\,\nabla\delta\chi\vb{\times}\ov{\mathbb{H}}\right) =  \pd{}{t}\left(\delta\chi\nabla\vb{\cdot}\ov{\mathbb{D}}\right) - \nabla\vb{\cdot}\left[ \delta\chi\left( \pd{\ov{\mathbb{D}}}{t} - c\,\nabla\vb{\times}\ov{\mathbb{H}}\right) \right], \]
 and, using the reduced Maxwell equations \eqref{eq:red_divD}-\eqref{eq:red_curlH} and the gauge-independent term $\delta\ov{\cal S} + (e/c)\,\delta\chi = m\ov{\bf v}\bdot\delta{\bf x} - \ov{K}\,\delta t$, the reduced Noether equation \eqref{eq:red_Noether} yields the reduced energy-momentum conservation law
 \begin{equation}
 \pd{}{t}\left(  \vb{\cal P}\bdot\delta{\bf x} \;-\frac{}{} {\cal E}\,\delta t \right) \;+\; \nabla\bdot\left(  {\sf T}\bdot\delta{\bf x} \;-\frac{}{} {\bf S}\,\delta t \right) \;=\; 0.
 \end{equation}
Here, the reduced energy-momentum densities
\begin{eqnarray}
{\cal E} & = & \int_{\ov{\bf p}} \ov{f}\,\ov{K} \;+\; \frac{1}{4\pi}\,{\bf E}\bdot\ov{\mathbb{D}} \;-\; \frac{1}{8\pi} \left( |{\bf E}|^{2} \;-\frac{}{} |{\bf B}|^{2} \right), \label{eq:red_E} \\
\vb{\cal P} & = & \int_{\ov{\bf p}} \ov{f}\,\frac{d\ov{\bf x}}{dt}\;m\,\ov{\bf v} \;+\; \frac{\ov{\mathbb{D}}\btimes{\bf B}}{4\pi\,c}. \label{eq:red_P}
\end{eqnarray}
both include reduced polarization effects, with Eq.~\eqref{eq:red_P} displaying the Minkowski form ($\ov{\mathbb{D}}\btimes{\bf B}/4\pi c$) for the reduced electromagnetic momentum density. The reduced energy-density flux
 \begin{equation}
 {\bf S} \;=\; \int_{\ov{\bf p}} \ov{f}\;\frac{d\ov{\bf x}}{dt}\,\ov{K} \;+\; \frac{c}{4\pi}\;{\bf E}\btimes\ov{\mathbb{H}},
 \label{eq:red_S}
 \end{equation}
 on the other hand, displays the Abraham form (${\bf E}\btimes c\,\ov{\mathbb{H}}/4\pi$) for the reduced Poynting flux, while the reduced stress tensor
 \begin{equation}
 {\sf T} \;=\; \int_{\ov{\bf p}} \ov{f}\;\frac{d\ov{\bf x}}{dt}\;m\,\ov{\bf v} + \frac{\bf I}{4\pi} \left[ \frac{1}{2} \left(|{\bf E}|^{2} \;-\; |{\bf B}|^{2}|\right) \;+\; {\bf B}\bdot\ov{\mathbb{H}} \right] - \frac{1}{4\pi} \left({\bf B}\,\ov{\mathbb{H}} \;+\frac{}{} \ov{\mathbb{D}}\,{\bf E}\right),
\label{eq:red_T}
\end{equation}
is composed of the reduced Reynolds stress tensor, which includes the ponderomotive velocity $\partial\ov{\Psi}/\partial\ov{\bf p} = d\ov{\bf x}/dt - \ov{\bf v}$, and the reduced Maxwell stress tensor, which includes polarization and magnetization corrections. We immediately see that the reduced stress tensor \eqref{eq:red_T}, which can be expressed as
\begin{eqnarray}
{\sf T} & = & \frac{\bf I}{8\pi} \left( |{\bf E}|^{2} + |{\bf B}|^{2} \right) \;-\; \frac{1}{4\pi} \left({\bf E}\,{\bf E} \;+\frac{}{} {\bf B}\,{\bf B} \right) \;+\; \int_{\ov{\bf p}} \ov{f}\; \left[ m\,\frac{d\ov{\bf x}}{dt}\;\frac{d\ov{\bf x}}{dt} \;+\; {\bf I}\;\left({\bf B}\bdot\pd{\ov{\Psi}}{\bf B}\right) \right] \nonumber \\
  &&-\; \int_{\ov{\bf p}} \ov{f} \left( m\,\frac{d\ov{\bf x}}{dt}\;\pd{\ov{\Psi}}{\ov{\bf p}} \;+\; {\bf B}\;\pd{\ov{\Psi}}{\bf B} \;-\; \pd{\ov{\Psi}}{\bf E}\;{\bf E} \right),
  \label{eq:red_stress}
\end{eqnarray} 
is manifestly asymmetric as a result of ponderomotive, polarization, and magnetization effects appearing on the last line of Eq.~\eqref{eq:red_stress}.

The apparent asymmetry of the reduced stress tensor \eqref{eq:red_stress} implies that the azimuthal angular momentum may not be conserved:
 \begin{equation}
 \pd{{\cal P}_{\varphi}}{t} \;+\; \nabla\bdot{\bf T}_{\varphi} \;=\; {\sf T}^{\top}:\nabla(\partial{\bf x}/\partial\varphi) \;\equiv\; \wh{\sf z}\bdot\left(\int_{\bf p} \ov{f}\;\ov{\bf N} \right),
 \label{eq:P_phi_red}
 \end{equation}
 unless the reduced torque $\ov{\bf N}$ vanishes identically. In Eq.~\eqref{eq:P_phi_red}, ${\cal P}_{\varphi} \equiv \vb{\cal P}\bdot\partial{\bf x}/\partial\varphi$ is the azimuthal angular momentum density, ${\bf T}_{\varphi} \equiv {\sf T}\bdot\partial{\bf x}/\partial\varphi$ is the azimuthal angular momentum-density flux, ${\sf T}^{\top}$ denotes the transpose of ${\sf T}$, and, since the dyadic tensor $\nabla(\partial{\bf x}/\partial\varphi)$ is antisymmetric, the reduced torque $\ov{\bf N}$ is expressed as
\begin{equation}
\ov{\bf N} \;\equiv\; \frac{d\ov{\bf x}}{dt}\btimes m\,\ov{\bf v} \;-\; \left(\ov{\vb{\pi}}\btimes{\bf E} \;+\frac{}{} \ov{\vb{\mu}}\btimes{\bf B}\right),
\label{eq:red_torque}
 \end{equation}
which includes contributions from the electric and magnetic torques. The required symmetry of the reduced stress tensor \eqref{eq:red_stress} must, therefore, introduce constraints on the reduced polarization and magnetization, which force the reduced torque \eqref{eq:red_torque} to vanish identically.
 
 \subsection{Guiding-center Vlasov-Maxwell equations}
 
The apparent asymmetry of the guiding-center stress tensor was independently noted by \cite{Pfirsch_Morrison_1985} and \cite{Similon_1985}, by using different variational formulations. It was recently shown by \cite{Brizard_Tronci_2016} and 
\cite{Sugama_2016}, however, that the guiding-center stress tensor is indeed explicitly symmetric. Here, we use the variational formulation of the guiding-center Vlasov-Maxwell model of \cite{Brizard_Tronci_2016} to show that the guiding-center torque, derived from the generic reduced torque \eqref{eq:red_torque}, vanishes identically.
 
 In the work of  \cite{Brizard_Tronci_2016}, which considers the simplest case ${\bf E} = 0$, the guiding-center canonical momentum is defined as $\ov{\bf p} = (e/c)\,{\bf A} + \ov{p}_{\|}\,\bhat$, which implies that $\ov{\bf v} = (\ov{p}_{\|}/m)\,\bhat$ and the guiding-center electric and magnetic dipole moments are
 \begin{equation}
 \left. \begin{array}{rcl}
 \ov{\vb{\pi}} &=& (e\bhat/\Omega)\btimes d\ov{\bf x}/dt \;=\; (e\bhat/\Omega)\btimes\partial\ov{\Psi}/\partial\ov{\bf p} \\
  && \\
\ov{\vb{\mu}} &=& -\,\ov{\mu}\,\bhat \;+\; \ov{\vb{\pi}}\btimes(\ov{p}_{\|}\bhat/mc)
\end{array} \right\}.
\label{eq:pimu_gc}
\end{equation}
Hence, using Eqs.~\eqref{eq:red_x} and \eqref{eq:pimu_gc}, we easily verify that the guiding-center torque \eqref{eq:red_torque} vanishes:
 \begin{equation}
\ov{\bf N} \;=\; \frac{d\ov{\bf x}}{dt}\btimes m\,\ov{\bf v} \;-\; \left( -\,\ov{\mu}\,\bhat + \ov{\vb{\pi}}\btimes\frac{\ov{p}_{\|}\bhat}{mc} \right)\btimes{\bf B} \;=\; \left(-\,m\,\ov{\bf v} \;+\; \ov{p}_{\|}\,\bhat\right) \btimes\frac{d\ov{\bf x}}{dt} \;\equiv\; 0,
\end{equation}
and the guiding-center stress tensor \eqref{eq:red_T} is symmetric \citep{Sugama_2016,Brizard_Tronci_2016}:
 \begin{eqnarray}
 {\sf T}_{\rm gc} &=& \frac{1}{4\pi}\left(\frac{\bf I}{2}\;|{\bf B}|^{2} -  {\bf B}\,{\bf B} \right) + \int_{\ov{\bf p}} \ov{f}\; \left[m\,\frac{d\ov{\bf x}}{dt}\;\left(\frac{d\ov{\bf x}}{dt} \;-\; \pd{\ov{\Psi}}{\ov{\bf p}}  \right) +  {\bf I}\;\left({\bf B}\bdot\pd{\ov{\Psi}}{\bf B}\right)
 \;-\; {\bf B}\;\pd{\ov{\Psi}}{\bf B} \right] \nonumber \\
  &=& \frac{1}{4\pi}\left(\frac{\bf I}{2}\;|{\bf B}|^{2} -  {\bf B}\,{\bf B} \right) + {\sf P}_{\rm CGL} + \int_{\ov{\bf p}} \ov{f}\; \left[ \ov{p}_{\|} \left( \pd{\ov{\Psi}}{\ov{\bf p}}\,\bhat + \bhat\, \pd{\ov{\Psi}}{\ov{\bf p}}  \right) \right],
  \label{eq:gc_T}
  \end{eqnarray}
  where ${\sf P}_{\rm CGL} = \int_{\ov{\bf p}}\ov{f}[(\ov{p}_{\|}^{2}/m)\bhat\bhat + \ov{\mu} B\,({\bf I} - \bhat\bhat)]$ is the symmetric Chew-Goldberger-Low (CGL) pressure tensor and the ponderomotive velocity $\partial\ov{\Psi}/\partial\ov{\bf p}$, which is assumed to be perpendicular to ${\bf B}$, represents the magnetic-drift velocity.
  
\section{\label{sec:symp_gyro}Gauge-free Gyrocenter Lagrangian Dynamics}

In this Section, we present two gauge-free gyrokinetic models whose gyrocenter equations of motion only involve the perturbed electromagnetic fields $({\bf E}_{1},{\bf B}_{1})$, thereby guaranteeing gauge freedom. Here, the separation of the perturbed magnetic field ${\bf B}_{1}$ from the unperturbed (equilibrium) magnetic field ${\bf B}_{0}$ satisfies the perturbed Faraday's Law $\partial{\bf B}_{1}/\partial t \equiv -\,c\,\nabla\btimes{\bf E}_{1}$, while the equilibrium magnetic field ${\bf B}_{0}$ is assumed to be toroidally axisymmetric, so that $\partial{\bf B}_{0}/\partial\varphi \equiv \wh{\sf z}\btimes{\bf B}_{0}$.

Gauge-free electromagnetic gyrokinetic Vlasov-Maxwell models were recently derived in the Hamiltonian representation by \cite{Burby_Brizard_2019} and in the symplectic representation by \cite{Brizard_2020}. The general form of the gauge-free gyrocenter Lagrangian is defined on gyrocenter phase space, with coordinates $({\bf X}, p_{\|}, \mu, \zeta)$, as
 \begin{equation}
 L_{\rm gy} = \left[ \frac{e}{c}\,\left( {\bf A}_{0}^{*} + \epsilon\,{\bf A}_{1{\rm gy}}\right) + \vb{\Pi}_{\rm gy}\right]\vb{\cdot}\dot{\bf X} + J\,\dot{\zeta} - \left( e\,\epsilon\,\Phi_{1{\rm gy}} \;+\frac{}{} K_{\rm gy} \right) \equiv {\bf P}_{\rm gy}\vb{\cdot}\dot{\bf X} + J\;\dot{\zeta} - H_{\rm gy},
 \label{eq:Lag_gy}
 \end{equation}
 where the the gyrocenter gyro-action $J \equiv (mc/e)\,\mu$ (which is canonically conjugate to the gyrocenter gyroangle $\zeta$) is used here only as a matter of convenience, and 
 \begin{equation}
 (e/c)\,{\bf A}_{0}^{*} \equiv (e/c)\,{\bf A}_{0} + p_{\|}\,\bhat_{0} - J\,({\bf R}_{0} + \frac{1}{2}\,\nabla\btimes\bhat_{0})
 \label{eq:A0_*}
 \end{equation} 
 is expressed in terms of the unperturbed (equilibrium) magnetic field ${\bf B}_{0} = \nabla\btimes{\bf A}_{0} = B_{0}\,\bhat_{0}$, and Eq.~\eqref{eq:A0_*} includes the gyrogauge vector field ${\bf R}_{0} \equiv \nabla\wh{\sf 1}\bdot\wh{\sf 2}$ (where $\bhat_{0} \equiv \wh{\sf 1}\btimes\wh{\sf 2}$) and higher-order guiding-center corrections \citep{Tronko_Brizard_2015} associated with the guiding-center electric-dipole moment $\vb{\pi}_{\rm gc} \equiv (e\bhat_{0}/\Omega_{0})\btimes\dot{\bf X}_{\rm gc}$. Similar higher-order guiding-center corrections are retained by \cite{Fan_2020}, with the main difference that, in our work, these guiding-center terms are explicitly time-independent (and non-variational).
  
\subsection{Gauge-free gyrocenter models}

In the gyrocenter Hamiltonian model of \cite{Burby_Brizard_2019}, presented here in the drift-kinetic limit considered by \cite{EH_2020}, we find the definitions
\begin{equation}
(\Phi_{1{\rm gy}},\; {\bf A}_{1{\rm gy}},\; \vb{\Pi}_{\rm gy}) \;=\; (\Phi_{1},\; {\bf A}_{1},\; 0),
\label{eq:PhiA_BB}
\end{equation}
where the perturbation fields are evaluated at the gyrocenter position ${\bf X}$, and the gyrocenter kinetic energy is
\begin{eqnarray}
K_{\rm gy} &=& \frac{p_{\|}^{2}}{2m} + \mu\,\left( B_{0} + \epsilon\,B_{1\|} + \frac{\epsilon^{2}}{2}\,\frac{|{\bf B}_{1}|^{2}}{B_{0}} \right) \;-\; \vb{\pi}_{\rm gc}\bdot\epsilon\left( {\bf E}_{1} + (p_{\|}\bhat_{0}/mc)\btimes
{\bf B}_{1}\right) \nonumber \\
 &&-\; \epsilon^{2}\,\frac{mc^{2}}{2B_{0}^{2}}\;\left| {\bf E}_{1} \;+\; (p_{\|}\bhat_{0}/mc)\btimes{\bf B}_{1}\right|^{2},
\label{eq:Kgy_BB}
\end{eqnarray}
where $B_{1\|} \equiv \bhat_{0}\bdot{\bf B}_{1}$ denotes the parallel component of the perturbed magnetic field ${\bf B}_{1}$. We note that the gauge-free model considered by \cite{EH_2020} omits the guiding-center electric-dipole moment $\vb{\pi}_{\rm gc}$ in the gyrocenter kinetic energy \eqref{eq:Kgy_BB} and, thus, the gyrocenter polarization and magnetization derived without this term are incomplete. We will explicitly show in Sec.~\ref{sec:symp_laws}, however, that this omission does not jeopardize the energy-momentum conservation laws.

Next, in the gyrocenter symplectic model of \cite{Brizard_2020}, we find
\begin{equation}
\left. \begin{array}{rcl}
(\Phi_{1{\rm gy}},\; {\bf A}_{1{\rm gy}}) & = & (\langle\Phi_{1{\rm gc}}\rangle,\; \langle{\bf A}_{1{\rm gc}}\rangle) \\
\vb{\Pi}_{\rm gy} & = & \epsilon\,\left(\langle{\bf E}_{1{\rm gc}}\rangle + (p_{\|}\bhat_{0}/mc)\btimes\langle{\bf B}_{1{\rm gc}}\rangle\right)\btimes e\bhat_{0}/\Omega_{0} 
\end{array} \right\},
\label{eq:PhiA_B}
\end{equation}
where perturbation fields are evaluated at ${\bf X} + \vb{\rho}_{0}$, with $\langle\cdots\rangle$ denoting the standard gyroangle averaging (since the lowest-order guiding-center gyroradius $\vb{\rho}_{0}$ depends on the gyrocenter gyroangle $\zeta$), and the gyrocenter kinetic energy is
\begin{eqnarray}
K_{\rm gy} &=& \frac{p_{\|}^{2}}{2m} + \mu\,\left( B_{0} + \epsilon\,\langle\langle B_{1\|{\rm gc}}\rangle\rangle + \frac{\epsilon^{2}}{2}\,\frac{|{\bf B}_{1}|^{2}}{B_{0}} \right) \nonumber \\
 &&+\; \epsilon^{2}\,\frac{mc^{2}}{2B_{0}^{2}}\;\left| {\bf E}_{1} \;+\; (p_{\|}\bhat_{0}/mc)\btimes{\bf B}_{1}\right|^{2}.
\label{eq:Kgy_B}
\end{eqnarray}
In Eq.~\eqref{eq:Kgy_B}, the finite-Larmor-radius (FLR) effects are included only at first order in the perturbation expansion, with $\langle\langle\cdots\rangle\rangle$ denoting the gyro-surface averaging introduced by \cite{Porazik_Lin_2011}. 

Previous symplectic gyrokinetic models considered either the parallel component $\langle A_{1\|{\rm gc}}\rangle$ of the perturbed vector potential \citep{HLB_1988,Brizard_2017b}, the inclusion of the perturbed $E\times B$ velocity \citep{Wang_Hahm_2010a,Wang_Hahm_2010b,Leerink_2010}, or both \citep{Duthoit_2014}. In the present symplectic gyrokinetic model \eqref{eq:PhiA_B}-\eqref{eq:Kgy_B}, the addition of the perturbed magnetic-flutter momentum to the $E\times B$ momentum yields a covariant treatment of the electric-dipole moment in the gyrocenter polarization and magnetization; see Eqs.~\eqref{eq:Pi_E}-\eqref{eq:KB_B}. In their guiding-center treatment, \cite{Fan_2020} considered an extension of the 
\cite{Pfirsch_Morrison_1985} variational formulation by including higher-order guiding-center corrections, where both electric and magnetic fields $({\bf E}, {\bf B} = B\,\bhat)$ are considered as variational fields.

\subsection{Gyrocenter Euler-Lagrange equations}

The gyrocenter Euler-Lagrange equations involving arbitrary variations in $({\bf X}, p_{\|}, J)$ are, respectively,
\begin{eqnarray}
0 & = & e\,{\bf E}_{\rm gy}^{*} \;+\; \frac{e}{c}\dot{\bf X}\btimes{\bf B}_{\rm gy}^{*} \;-\; \dot{p}_{\|}\;{\sf b}_{\rm gy}^{*}, \label{eq:EL_X} \\
0 & = & \dot{\bf X}\bdot{\sf b}_{\rm gy}^{*} \;-\; \partial K_{\rm gy}/\partial p_{\|}, \label{eq:EL_p} \\
0 & = & \dot{\zeta} \;+\; \dot{\bf X}\bdot\partial{\bf P}_{\rm gy}/\partial J \;-\; \partial H_{\rm gy}/\partial J,
\end{eqnarray}
where the effective gyrocenter electric field ${\bf E}_{\rm gy}^{*}$ is defined as
\begin{equation}
e\,{\bf E}_{\rm gy}^{*} \;\equiv\; -\,\nabla H_{\rm gy} - \pd{{\bf P}_{\rm gy}}{t} \;=\; \epsilon\,e\,{\bf E}_{1{\rm gy}} \;-\; \left( \pd{\vb{\Pi}_{\rm gy}}{t} \;+\; \nabla K_{\rm gy} \right),
 \label{eq:Egy_*}
\end{equation}
with ${\bf E}_{1{\rm gy}} \equiv -\,\nabla\Phi_{1{\rm gy}} - c^{-1}\partial{\bf A}_{1{\rm gy}}/\partial t$, and the effective gyrocenter magnetic field ${\bf B}_{\rm gy}^{*}$ is defined as
\begin{equation}
{\bf B}_{\rm gy}^{*} \;\equiv\; \nabla\btimes\left(\frac{c}{e}\,{\bf P}_{\rm gy}\right) \;=\; {\bf B}_{0}^{*} + \epsilon\,{\bf B}_{1{\rm gy}} \;+\;  \nabla\btimes\left(\frac{c}{e}\,\vb{\Pi}_{\rm gy}\right),
  \label{eq:Bgy_*}
\end{equation}
with ${\bf B}_{0}^{*} \equiv \nabla\btimes{\bf A}_{0}^{*}$ and ${\bf B}_{1{\rm gy}} \equiv \nabla\btimes{\bf A}_{1{\rm gy}}$, while
\begin{equation}
{\sf b}^{*}_{\rm gy} \;\equiv\; \partial{\bf P}_{\rm gy}/\partial p_{\|} \;=\; \bhat_{0} \;+\; \partial\vb{\Pi}_{\rm gy}/\partial p_{\|}.
\end{equation}
We note that the effective gyrocenter electromagnetic fields satisfy the source-free Maxwell equations $\nabla\bdot{\bf B}_{\rm gy}^{*} = 0$ and 
$\partial{\bf B}_{\rm gy}^{*}/\partial t + c\,\nabla\btimes{\bf E}_{\rm gy}^{*} = 0$.

The gyrocenter Euler-Lagrange equations \eqref{eq:EL_X}-\eqref{eq:EL_p} can also be written in Hamiltonian form as
\begin{eqnarray}
\dot{\bf X}  & \equiv & \left\{ {\bf X},\; H_{\rm gy}\right\}_{\rm gy} \;=\; {\bf E}_{\rm gy}^{*}\btimes\frac{c{\sf b}^{*}_{\rm gy}}{B_{\|{\rm gy}}^{**}} + \pd{K_{\rm gy}}{p_{\|}}\,\frac{{\bf B}^{*}_{\rm gy}}{B_{\|{\rm gy}}^{**}}, \label{eq:Xgy_dot} \\
\dot{p}_{\|} & \equiv & \left\{ p_{\|},\; H_{\rm gy}\right\}_{\rm gy} \;=\; e\,{\bf E}_{\rm gy}^{*}\bdot\frac{{\bf B}^{*}_{\rm gy}}{B_{\|{\rm gy}}^{**}}, \label{eq:pgy_dot}
\end{eqnarray}
where $\{\;,\;\}_{\rm gy}$ denotes the gyrocenter Poisson bracket and $B_{\|{\rm gy}}^{**} \equiv {\sf b}^{*}_{\rm gy}\bdot{\bf B}^{*}_{\rm gy}$. We note that Eqs.~\eqref{eq:Xgy_dot}-\eqref{eq:pgy_dot} satisfy the Euler-Lagrange identity:
\begin{equation}
\pd{K_{\rm gy}}{p_{\|}}\,\dot{p}_{\|} \;=\; e\,{\bf E}_{\rm gy}^{*}\bdot\pd{K_{\rm gy}}{p_{\|}}\,\frac{{\bf B}^{*}_{\rm gy}}{B_{\|{\rm gy}}^{**}} \;\equiv\; e\,{\bf E}_{\rm gy}^{*}\bdot\dot{\bf X},
\label{eq:xp_dot_id}
\end{equation}
which will be useful in our discussion of energy conservation. The gyrocenter equations \eqref{eq:Xgy_dot}-\eqref{eq:pgy_dot} also satisfy the Liouville Theorem
\begin{eqnarray}
\pd{B_{\|{\rm gy}}^{**}}{t} & = & \pd{{\sf b}_{\rm gy}^{*}}{t}\bdot{\bf B}_{\rm gy}^{*} \;+\; {\sf b}_{\rm gy}^{*}\bdot\pd{{\bf B}_{\rm gy}^{*}}{t} = \pd{}{p_{\|}}\left(\pd{{\bf P}_{\rm gy}}{t}\right)\bdot{\bf B}_{\rm gy}^{*} \;-\;
{\sf b}_{\rm gy}^{*}\bdot\nabla\btimes\left(c\,{\bf E}_{\rm gy}^{*}\right) \nonumber \\
 &=& -\pd{}{p_{\|}}\left[\nabla\bdot\left(H_{\rm gy}\,{\bf B}^{*}_{\rm gy}\right) \;+\; \dot{p}_{\|}\;B_{\|{\rm gy}}^{**}\right] \;-\; \pd{{\bf P}_{\rm gy}}{t}\bdot\nabla\btimes\left(\frac{c{\sf b}^{*}_{\rm gy}}{e}\right) \nonumber \\
  &&-\; \nabla\bdot\left[ B_{\|{\rm gy}}^{**}\;\dot{\bf X} \;-\; \pd{}{p_{\|}}\left( H_{\rm gy}\,{\bf B}^{*}_{\rm gy}\right) \right] \;-\; \left( e\,{\bf E}_{\rm gy}^{*} \;+\; \nabla H_{\rm gy}\right)\bdot\nabla\btimes\left(\frac{c{\sf b}^{*}_{\rm gy}}{e}\right) \nonumber \\
 & = & -\;\nabla\bdot\left(B_{\|{\rm gy}}^{**}\;\dot{\bf X}\right) \;-\; \pd{}{p_{\|}} \left( B_{\|{\rm gy}}^{**}\;\dot{p}_{\|} \right),
\label{eq:Liouville}
\end{eqnarray}
where we used Eq.~\eqref{eq:Egy_*}.

\subsection{Eulerian field variations of the gyrocenter Lagrangian}

In the next Section, we will need the Eulerian field variation of the gyrocenter Lagrangian \eqref{eq:Lag_gy} at a field point ${\bf x}$:
\begin{equation}
 \delta L_{\rm gy} = \left( \frac{e}{c}\,\epsilon\,\delta{\bf A}_{1{\rm gy}} + \delta\vb{\Pi}_{\rm gy}\right)\vb{\cdot}\dot{\bf X} - \left( e\,\epsilon\,\delta\Phi_{1{\rm gy}} \;+\frac{}{} \delta K_{\rm gy} \right),
 \label{eq:Lag_gy_var}
 \end{equation}
 where, in contrast to the works of \cite{Sugama_2021} and \cite{Fan_2020}, the guiding-center Lagrangian terms $(e/c){\bf A}_{0}^{*}\bdot\dot{\bf X} + J\dot{\zeta} - (p_{\|}^{2}/2m + \mu B_{0})$ are invariant in our gyrokinetic formalism. While \cite{Sugama_2021} considered the simplest guiding-center representation (with $e{\bf A}_{0}^{*}/c = e{\bf A}_{0}/c + p_{\|}\bhat$), with field variations easily computed (e.g., $\delta\bhat = (\bhat\btimes\delta{\bf B})\btimes\bhat/B$), the higher-order guiding-center model used by \cite{Fan_2020} requires complex expressions for the variations of the gyrogauge vector ${\bf R}_{0} = \nabla\wh{\sf 1}\bdot\wh{\sf 2}$, for example, in which the functional derivatives of all three unit vectors $(\wh{\sf 1},\wh{\sf 2}, \bhat = \wh{\sf 1}\btimes\wh{\sf 2})$ need to be computed, although they are not explicitly calculated.
 
 Here, the field variations are defined in terms of the generic functional derivatives
 \begin{equation}
\left( \begin{array}{c}
\delta\Psi_{1}({\bf X}) \\
\langle\delta\Psi_{1}({\bf X} + \vb{\rho}_{0})\rangle \end{array} \right) \;\equiv\; \int_{\bf x} \delta\Psi_{1}({\bf x})\;\left( \begin{array}{c}
\delta^{3}({\bf X} - {\bf x}) \\
\left\langle\delta^{3}({\bf X} + \vb{\rho}_{0} - {\bf x})\right\rangle
\end{array} \right),
\label{eq:delta_gy}
\end{equation}
where $\Psi_{1}$ denotes an arbitrary component of the perturbed electromagnetic potentials or fields. We note that the second expression in Eq.~\eqref{eq:delta_gy} is valid if only the equilibrium (non-variational) magnetic field appears in the definition of the lowest-order gyroangle-dependent gyroradius $\vb{\rho}_{0}$. Hence, we find
 \begin{equation}
 \left(\fd{\Phi_{1{\rm gy}}}{\Phi_{1}({\bf x})}, \fd{A^{i}_{1{\rm gy}}}{A^{j}_{1}({\bf x})}, \fd{E^{i}_{1{\rm gy}}}{E^{j}_{1}({\bf x})}, \fd{B^{i}_{1{\rm gy}}}{B^{j}_{1}({\bf x})} \right) \;=\; \left\{ \begin{array}{l}
 \left( \delta^{3},\; \delta^{i}_{j}\,\delta^{3},\; \delta^{i}_{j}\,\delta^{3},\; \delta^{i}_{j}\,\delta^{3}\right) \\
 \\
 \left( \langle\delta_{\rm gc}^{3}\rangle,\; \delta^{i}_{j}\,\langle\delta_{\rm gc}^{3}\rangle,\; \delta^{i}_{j}\,\langle\delta_{\rm gc}^{3}\rangle,\; \delta^{i}_{j}\,\langle\delta_{\rm gc}^{3}\rangle\right)
 \end{array} \right. 
 \label{eq:delta_defs}
 \end{equation}
 and
  \[ \left( \fd{B_{1\|}({\bf X})}{{\bf B}_{1}({\bf x})},\; \fd{\langle\langle B_{1\|{\rm gc}}\rangle\rangle}{{\bf B}_{1}({\bf x})}\right) \;=\; \left( \delta^{3}\,\bhat_{0},\; \langle\langle\delta_{\rm gc}^{3}\rangle\rangle\,\bhat_{0}\right), \]
 with $\delta^{3} \equiv \delta^{3}({\bf X} - {\bf x})$ and $\delta_{\rm gc}^{3} \equiv \delta^{3}({\bf X} + \vb{\rho}_{0} - {\bf x})$ used in the gyrocenter models of \cite{Burby_Brizard_2019} and \cite{Brizard_2020}, respectively, and $\delta^{i}_{j}$ denotes the standard Kronecker delta. 
 
 In the gyrocenter model \eqref{eq:Kgy_BB} of \cite{Burby_Brizard_2019}, we find
 \begin{eqnarray}
\epsilon^{-1} \fd{K_{\rm gy}}{{\bf E}_{1}({\bf x})} & = & -\,\delta^{3} \left[ \vb{\pi}_{\rm gc} \;+\; \epsilon\,\frac{mc^{2}}{B_{0}^{2}} \left({\bf E}_{1} + \frac{p_{\|}\bhat_{0}}{mc}\btimes{\bf B}_{1}\right) \right] \equiv -\,\delta^{3}\,\left(\vb{\pi}_{\rm gc} + 
\epsilon\,\vb{\pi}_{2}\right), 
 \label{eq:KE_BB} \\
\epsilon^{-1}  \fd{K_{\rm gy}}{{\bf B}_{1}({\bf x})} & = & \delta^{3}\,\mu \left( \bhat_{0} + \epsilon\,\frac{{\bf B}_{1}}{B_{0}}\right) - \delta^{3}\left(\vb{\pi}_{\rm gc} \;+\; \epsilon\,\vb{\pi}_{2}\right)\btimes\frac{p_{\|}\bhat_{0}}{mc},
 \label{eq:KB_BB}
 \end{eqnarray}
 where the gyrocenter electric-dipole moment $\vb{\pi}_{\rm gc} + \epsilon\,\vb{\pi}_{2}$ includes the guiding-center contribution $\vb{\pi}_{\rm gc}$ and its first-order gyrocenter correction $\vb{\pi}_{2}$ (derived from the second-order gyrocenter Hamiltonian), while the intrinsic gyrocenter magnetic-dipole moment $-\,\mu(\bhat_{0} + \epsilon\, {\bf B}_{1}/B_{0})$ is accompanied by the moving gyrocenter electric-dipole moment contribution $(\vb{\pi}_{\rm gc} + \epsilon\,\vb{\pi}_{2})\btimes p_{\|}\bhat_{0}/mc$.
 
In the gyrocenter model \eqref{eq:PhiA_B}-\eqref{eq:Kgy_B} of \cite{Brizard_2020}, on the other hand, we find
 \begin{eqnarray}
 \epsilon^{-1} \fd{\vb{\Pi}_{\rm gy}}{{\bf E}_{1}({\bf x})}\bdot\dot{\bf X} & = & \langle\delta_{\rm gc}^{3}\rangle\;\frac{e\bhat_{0}}{\Omega_{0}}\btimes\dot{\bf X} \;\equiv\; \langle\delta_{\rm gc}^{3}\rangle\;\vb{\pi}_{\rm gy}.
 \label{eq:Pi_E}  \\
 \epsilon^{-1} \fd{\vb{\Pi}_{\rm gy}}{{\bf B}_{1}({\bf x})}\bdot\dot{\bf X} & = & \langle\delta_{\rm gc}^{3}\rangle\left(\vb{\pi}_{\rm gy} \btimes\frac{p_{\|}\bhat_{0}}{mc}\right),
 \label{eq:Pi_B}
 \end{eqnarray} 
and
 \begin{eqnarray}
 \epsilon^{-1} \fd{K_{\rm gy}}{{\bf E}_{1}({\bf x})} & = &  \epsilon\,\delta^{3}\,\frac{mc^{2}}{B_{0}^{2}} \left({\bf E}_{1} + \frac{p_{\|}\bhat_{0}}{mc}\btimes{\bf B}_{1}\right) \;\equiv\; \epsilon\,\delta^{3}\,\vb{\pi}_{2}, 
 \label{eq:KE_B} \\
 \epsilon^{-1} \fd{K_{\rm gy}}{{\bf B}_{1}({\bf x})} & = & \mu \left( \langle\langle\delta_{\rm gc}^{3}\rangle\rangle\,\bhat_{0} + \epsilon\,\delta^{3}\,{\bf B}_{1}/B_{0}\right) \;+\; \epsilon\,\delta^{3}\left( \vb{\pi}_{2}\btimes\frac{p_{\|}\bhat_{0}}{mc} \right).
 \label{eq:KB_B}
 \end{eqnarray} 
 
 We note that the gyrocenter polarization and magnetization derived from Eq.~\eqref{eq:KE_BB}-\eqref{eq:KB_BB} for the gyrokinetic model of \cite{Burby_Brizard_2019} are explicitly truncated at first order in the perturbation amplitudes of the electric and
magnetic fields $({\bf E}_{1},{\bf B}_{1})$. Because the gyrocenter velocity \eqref{eq:Xgy_dot} appears in the expressions \eqref{eq:Pi_E}-\eqref{eq:Pi_B} for the gyrokinetic model of \cite{Brizard_2020}, however, the corresponding gyrocenter polarization and magnetization contain contributions at higher orders in perturbation amplitude.
  
\section{\label{sec:symp_VP}Gyrokinetic Variational Principle}

The gyrokinetic Vlasov-Maxwell equations can be derived from several equivalent variational principles: Low-Lagrange \citep{Sugama_2000}; Euler \citep{Brizard_PRL_2000,Brizard_PoP_2000,Brizard_2009,Brizard_2010a,Brizard_2017b}; Hamilton-Jacobi \citep{CRP_2004}; and Euler-Poincar\'{e} \citep{Squire_2013,EH_2020}.  In recent work, \cite{Brizard_Tronci_2016} showed how the guiding-center Vlasov-Maxwell equations (derived without a separation between time-independent equilibrium and variational dynamical plasma fields) can be explicitly derived from many if these equivalent variational principles. 

In the present work, the separation of equilibrium and perturbed electromagnetic fields introduces a low-frequency gyrokinetic space-time ordering that assumes that the nonuniform equilibrium magnetic field is time-independent and non-variational. Applications of Noether's Theorem, which will explicitly take into account the properties of the equilibrium magnetic field, follow most naturally from an Eulerian variational principle. In recent work, \cite{EH_2020} derived the energy-momentum and angular-momentum conservation laws (without proof) within an Euler-Poincar\'{e} variational formulation for the Vlasov-Maxwell and drift-kinetic Vlasov-Maxwell equations described by the gauge-free gyrocenter model of \cite{Burby_Brizard_2019}.

We are now ready to derive the gauge-free gyrokinetic Vlasov-Maxwell equations from an Eulerian variational principle $\delta{\cal A}_{\rm gy} = 0$, based on the gyrokinetic action functional  \citep{Brizard_PoP_2000}
\begin{equation}
{\cal A}_{\rm gy} \equiv -\,\int {\cal F}_{\rm gy}\,{\cal H}_{\rm gy}\,d^{8}{\cal Z} + \int\frac{d^{4}x}{8\pi} \left( |{\bf E}|^{2} \;-\frac{}{} |{\bf B}|^{2}\right),
\label{eq:A_gy}
\end{equation}
where summation over particle species is implicitly assumed in the first term and the infinitesimal extended phase-space volume element $d^{8}{\cal Z}$ does not include the Jacobian ${\cal J}_{\rm gy}$. Instead, the perturbation-field-dependent Jacobian is inserted in the definition of the gyrocenter extended Vlasov density 
\begin{equation}
{\cal F}_{\rm gy} \;\equiv\; {\cal J}_{\rm gy}\,{\cal F} \;\equiv\; {\cal J}_{\rm gy}\,F\,\delta(w - H_{\rm gy}),
\end{equation}
which also includes an energy delta function that enforces the constraint ${\cal H}_{\rm gy} = H_{\rm gy} - w \equiv 0$ in extended gyrocenter phase space.

The variation of the gyrokinetic action functional yields
\begin{equation}
\delta{\cal A}_{\rm gy} \;=\; -\,\int \left(\delta{\cal F}_{\rm gy}\,{\cal H}_{\rm gy} \;+\frac{}{} {\cal F}_{\rm gy}\,\delta{\cal H}_{\rm gy}\right)\,d^{8}{\cal Z} \;+\; \int\frac{d^{4}x}{4\pi} \left( \epsilon\,\delta{\bf E}_{1}\bdot{\bf E} \;-\frac{}{} \epsilon\,
\delta{\bf B}_{1}\bdot{\bf B}\right),
 \label{eq:delta_Lgy}
\end{equation}
where the constrained electromagnetic variations 
\begin{equation}
\left. \begin{array}{rcl}
\delta{\bf E}_{1}({\bf x}) &\equiv& -\,\nabla\delta\Phi_{1}({\bf x}) \;-\; c^{-1}\partial_{t}\delta {\bf A}_{1}({\bf x}) \\
\delta{\bf B}_{1}({\bf x}) &\equiv& \nabla\btimes\delta {\bf A}_{1}({\bf x})
\end{array} \right\}
\label{eq:delta_EB}
\end{equation}
satisfy the Faraday constraint equation $\nabla\delta{\bf E}_{1} + c^{-1}\partial\delta{\bf B}_{1}/\partial t = 0$ and $\nabla\bdot\delta{\bf B}_{1} = 0$, with the equilibrium magnetic field ${\bf B}_{0}$ held constant under field variations. The variation of the gyrocenter Hamiltonian
\begin{equation}
\delta{\cal H}_{\rm gy} \;=\; \epsilon\,e\;\delta\Phi_{1{\rm gy}} \;+\; \delta{\bf E}_{1}\bdot\fd{K_{\rm gy}}{{\bf E}_{1}} \;+\; \delta{\bf B}_{1}\bdot\fd{K_{\rm gy}}{{\bf B}_{1}}
\label{eq:delta_Hgy}
\end{equation}
is expressed in terms of $\delta\Phi_{1}$ and $(\delta{\bf E}_{1},\delta{\bf B}_{1})$. The variation of the gyrocenter extended Vlasov density $\delta{\cal F}_{\rm gy} \equiv \delta{\cal J}_{\rm gy}\;{\cal F} + {\cal J}_{\rm gy}\;\delta{\cal F}$ is expressed as
\begin{eqnarray}
\delta{\cal F}_{\rm gy} &=& {\cal F} \left( \pd{\delta{\bf P}_{\rm gy}}{p_{\|}}\bdot\frac{e}{c}{\bf B}^{*}_{\rm gy} + {\sf b}^{*}_{\rm gy}\bdot\nabla\btimes\delta{\bf P}_{\rm gy}\right) + {\cal J}_{\rm gy} \left( \{ \delta{\cal S},\; {\cal F}\}_{\rm gy} \;+\frac{}{} 
\delta{\bf P}_{\rm gy}\bdot\{{\bf X},\; {\cal F}\}_{\rm gy} \right) \nonumber \\
 &\equiv& -\;\pd{}{{\cal Z}^{a}}\left( \delta{\cal Z}^{a}\frac{}{}{\cal F}_{\rm gy}\right),
 \label{eq:delta_Fgy}
\end{eqnarray}
where the virtual extended phase-space displacement
\begin{equation}
\delta{\cal Z}^{a} \;\equiv\; \left\{ {\cal Z}^{a},\frac{}{} \delta{\cal S}\right\}_{\rm gy} \;-\; \delta{\bf P}_{\rm gy}\bdot\left\{ {\bf X},\frac{}{} {\cal Z}^{a}\right\}_{\rm gy}
\end{equation}
is defined in terms of a canonical part generated by $\delta{\cal S}$ and a non-canonical part generated by 
\begin{equation}
\delta{\bf P}_{\rm gy} = \epsilon\,\frac{e}{c}\,\delta{\bf A}_{1{\rm gy}} \;+\; \delta{\bf E}_{1}\bdot\pd{\vb{\Pi}_{\rm gy}}{{\bf E}_{1}} \;+\; \delta{\bf B}_{1}\bdot\pd{\vb{\Pi}_{\rm gy}}{{\bf B}_{1}}.
\label{eq:delta_Pi_gy}
\end{equation}
We note that the final form in Eq.~\eqref{eq:delta_Fgy} for the Eulerian variation $\delta{\cal F}_{\rm gy}$ is a natural phase-space generalization of the variation $\delta n = -\,\nabla\bdot(\delta{\bf x}\,n)$ for the fluid particle density $n$. In addition, in the Hamiltonian model of \cite{Burby_Brizard_2019}, the last two terms are absent.

The first two variations in the Vlasov term in Eq.~\eqref{eq:delta_Lgy} can be combined
\begin{eqnarray}
-\,\delta({\cal F}_{\rm gy}\,{\cal H}_{\rm gy}) &=& -\, {\cal J}_{\rm gy}\{ {\cal F},\; {\cal H}_{\rm gy}\}_{\rm gy}\;\delta{\cal S} \;+\; {\cal F}_{\rm gy}\;\delta L_{\rm gy} \nonumber \\
 &&+\; \pd{}{t}\left({\cal F}_{\rm gy}\,\delta{\cal S}\right) \;+\; \nabla\bdot\left(\dot{\bf X}\;{\cal F}_{\rm gy}\,\delta{\cal S}\right) \;+\; \pd{}{p_{\|}}\left( \dot{p}_{\|}\;{\cal F}_{\rm gy}\,\delta{\cal S}\right),
\label{eq:delta_FH}
\end{eqnarray}
where the variation of the gyrocenter Lagrangian \eqref{eq:Lag_gy} is
\begin{eqnarray}
\delta L_{\rm gy} & \equiv & \epsilon\left( \frac{e}{c}\,\delta{\bf A}_{1{\rm gy}}\bdot\dot{\bf X} \;-\; e\,\delta\Phi_{1{\rm gy}} \right) \;+\; \delta{\bf E}_{1}\bdot\left( \pd{\vb{\Pi}_{\rm gy}}{{\bf E}_{1}}\bdot\dot{\bf X} \;-\; \pd{K_{\rm gy}}{{\bf E}_{1}} \right) 
\nonumber \\
 &&+\; \delta{\bf B}_{1}\bdot\left( \pd{\vb{\Pi}_{\rm gy}}{{\bf B}_{1}}\bdot\dot{\bf X} \;-\; \pd{K_{\rm gy}}{{\bf B}_{1}} \right).
 \end{eqnarray}
Using Eq.~\eqref{eq:delta_defs} and Eqs.~\eqref{eq:KE_BB}-\eqref{eq:KB_BB} or \eqref{eq:Pi_E}-\eqref{eq:KB_B}, the Lagrangian variation term
 \begin{equation}
 \int_{\cal Z} {\cal F}_{\rm gy}\;\delta L_{\rm gy} \;=\; \int_{\bf x} \left( \frac{\epsilon}{c}\,\delta{\bf A}_{1}\bdot{\bf J}_{\rm gy} \;-\; \epsilon\,\delta\Phi_{1}\;\varrho_{\rm gy} \;+\; \epsilon\,\delta{\bf E}_{1}\bdot\mathbb{P}_{\rm gy} + \epsilon\,
 \delta{\bf B}_{1}\bdot\mathbb{M}_{\rm gy} \right)
\label{eq:Pi_Psi}
\end{equation}
 can be expressed in terms of the gyrocenter charge and current densities
\begin{eqnarray}
\left(\varrho_{\rm gy}({\bf x}),\frac{}{} {\bf J}_{\rm gy}({\bf x})\right) & \equiv & \int_{\cal Z} {\cal F}_{\rm gy}\; \left( -\,\epsilon^{-1}\fd{L_{\rm gy}}{\Phi_{1}({\bf x})},\; \epsilon^{-1}\fd{L_{\rm gy}}{{\bf A}_{1}({\bf x})}\right) \nonumber \\
 &=& \int_{\cal Z} {\cal F}_{\rm gy}\; \left( e\,\fd{\Phi_{1{\rm gy}}}{\Phi_{1}({\bf x})},\; e\,\fd{{\bf A}_{1{\rm gy}}}{{\bf A}_{1}({\bf x})}\bdot\dot{\bf X}\right),
\label{eq:rhoJ_gy}
\end{eqnarray}
and the gyrocenter polarization and magnetization
\begin{eqnarray}
\mathbb{P}_{\rm gy}({\bf x}) & \equiv &  \int_{\cal Z} {\cal F}_{\rm gy}\;\epsilon^{-1}\fd{L_{\rm gy}}{{\bf E}_{1}({\bf x})} \;=\; \int_{\cal Z} {\cal F}_{\rm gy} \left( \epsilon^{-1}\fd{\vb{\Pi}_{\rm gy}}{{\bf E}_{1}({\bf x})}\bdot\dot{\bf X} \;-\; 
\epsilon^{-1}\fd{K_{\rm gy}}{{\bf E}_{1}({\bf x})} \right), \label{eq:Pol_gy} \\
\mathbb{M}_{\rm gy}({\bf x})  & \equiv &  \int_{\cal Z} {\cal F}_{\rm gy}\;\epsilon^{-1}\fd{L_{\rm gy}}{{\bf B}_{1}({\bf x})} \;=\;  \int_{\cal Z} {\cal F}_{\rm gy} \left( \epsilon^{-1}\fd{\vb{\Pi}_{\rm gy}}{{\bf B}_{1}({\bf x})}\bdot\dot{\bf X} \;-\; 
\epsilon^{-1}\fd{K_{\rm gy}}{{\bf B}_{1}({\bf x})} \right).
\label{eq:Mag_gy}
\end{eqnarray}
When the gauge-free gyrokinetic models represented by Eqs.~\eqref{eq:PhiA_BB}-\eqref{eq:Kgy_BB} and \eqref{eq:PhiA_B}-\eqref{eq:Kgy_B} are used, the gyrocenter polarization is given for the \cite{Burby_Brizard_2019} model (top) and the \cite{Brizard_2020} model (bottom) as
\begin{equation}
\mathbb{P}_{\rm gy}({\bf x}) \;=\;  \int_{\cal Z} {\cal F}_{\rm gy}\;\left\{ \begin{array}{l}
\delta^{3}\,\left(\vb{\pi}_{\rm gc} \;+\; \epsilon\,\vb{\pi}_{2}\right)\\
\\
\langle\delta_{\rm gc}^{3}\rangle\,\vb{\pi}_{\rm gy} - \epsilon\,\delta^{3}\,\vb{\pi}_{2}
 \end{array} \right.
 \end{equation} 
  where $\vb{\pi}_{2}$ and $\vb{\pi}_{\rm gy}$ are defined in Eqs.~\eqref{eq:KE_BB} and \eqref{eq:Pi_E}, respectively, and the gyrocenter magnetization is
\begin{equation}
\mathbb{M}_{\rm gy}({\bf x}) \;=\;  \int_{\cal Z} {\cal F}_{\rm gy}\;\left\{ \begin{array}{l}
\delta^{3}\,\left[ -\,\mu\;\left(\bhat_{0} + \epsilon\,{\bf B}_{1}/B_{0}\right)  \;+\; \left(\vb{\pi}_{\rm gc} \;+\; \epsilon\,\vb{\pi}_{2}\right)\btimes(p_{\|}\bhat_{0}/mc)\right] \\
\\
-\,\mu \left(\langle\langle\delta_{\rm gc}^{3}\rangle\rangle\,\bhat_{0} + \epsilon\,\delta^{3}\,{\bf B}_{1}/B_{0}\right) \\
+\; \left(\langle\delta_{\rm gc}^{3}\rangle\,\vb{\pi}_{\rm gy} - \epsilon\,\delta^{3}\,\vb{\pi}_{2}\right)\btimes(p_{\|}\bhat_{0}/mc)
 \end{array} \right.
 \end{equation} 
 We note, here, that the lowest-order guiding-center contributions to polarization and magnetization are derived from the first-order gyrocenter Lagrangian, which circumvents the need to consider guiding-center variations as in the works of \cite{Sugama_2021} and \cite{Fan_2020}.
 
The variation of the Maxwell Lagrangian density can be expressed as 
\begin{eqnarray}
\delta{\bf E}_{1}\bdot{\bf E} - \delta{\bf B}_{1}\bdot{\bf B} &=& \delta {\bf A}_{1}\bdot\left(\frac{1}{c}\pd{\bf E}{t} - \nabla\btimes{\bf B}\right) \;+\; \delta\Phi_{1}\;(\nabla\bdot{\bf E}) \nonumber \\
 &&-\; \pd{}{t}\left(\frac{1}{c}\,\delta {\bf A}_{1}\bdot{\bf E}\right) \;-\; \nabla\bdot\left(\delta\Phi_{1}\,{\bf E} \;+\frac{}{} \delta {\bf A}_{1}\btimes{\bf B}\right).
\label{eq:delta_EB_energy}
\end{eqnarray}
If we now combine Eqs.~\eqref{eq:delta_FH}-\eqref{eq:delta_EB_energy} into the variation of the gyrokinetic action functional \eqref{eq:delta_Lgy}: $\delta{\cal A}_{\rm gy} \equiv \int \delta{\cal L}_{\rm gy}\,d^{4}x$, we obtain the variation of the gyrokinetic Lagrangian density
\begin{eqnarray}
\delta{\cal L}_{\rm gy} & = & -\,\int_{P} {\cal J}_{\rm gy}\{ {\cal F},\; {\cal H}_{\rm gy}\}_{\rm gy}\;\delta{\cal S} \;+\; \frac{\epsilon\,\delta\Phi_{1}}{4\pi} \left( \nabla\bdot\mathbb{D}_{\rm gy} \;-\frac{}{} 4\pi \;\varrho_{\rm gy}  \right) \nonumber \\
 &&+\; \frac{\epsilon}{4\pi}\,\delta {\bf A}_{1}\bdot\left( \frac{1}{c}\pd{\mathbb{D}_{\rm gy}}{t} - \nabla\btimes\mathbb{H}_{\rm gy} + \frac{4\pi}{c}\;{\bf J}_{\rm gy}  \right) \nonumber \\
  &  &+\; \pd{}{t}\left(\int_{P}{\cal F}_{\rm gy}\,\delta{\cal S} - \frac{\epsilon}{4\pi}\,\delta {\bf A}_{1}\bdot\mathbb{D}_{\rm gy}\right) \nonumber \\
  &&+\; \nabla\bdot\left(\int_{P}\dot{\bf X}{\cal F}_{\rm gy}\delta{\cal S} - \frac{\epsilon}{4\pi}\left(\delta\Phi_{1}\mathbb{D}_{\rm gy} + \delta {\bf A}_{1}\times\mathbb{H}_{\rm gy}\right) \right),
  \label{eq:Lag_gy_vp}
\end{eqnarray}
where the gyrocenter macroscopic electromagnetic fields are defined as
\begin{equation}
\left. \begin{array}{rcl}
\mathbb{D}_{\rm gy} & \equiv & \epsilon\,{\bf E}_{1} \;+\; 4\pi\,\mathbb{P}_{\rm gy} \\
\mathbb{H}_{\rm gy} & \equiv & {\bf B}_{0} \;+\; \epsilon\,{\bf B}_{1} \;-\; 4\pi\,\mathbb{M}_{\rm gy}
\end{array} \right\},
\label{eq:DH_def}
\end{equation}
and the variations $(\delta{\cal S}, \delta\Phi_{1}, \delta{\bf A}_{1})$ are assumed to be arbitrary. Variation with respect to $\delta{\cal S}$ yields the gyrokinetic Vlasov equation in extended phase space $\{ {\cal F},\; 
{\cal H}_{\rm gy}\}_{\rm gy} = 0$. If we integrate ${\cal J}_{\rm gy}\{ {\cal F},\; {\cal H}_{\rm gy}\}_{\rm gy}$ over the energy $w$ coordinate, we find
\begin{eqnarray}
0 & = & \int {\cal J}_{\rm gy}\{ {\cal F},\; {\cal H}_{\rm gy}\}_{\rm gy}\;dw \;=\; \int \pd{}{{\cal Z}^{a}}\left({\cal J}_{\rm gy}\,{\cal F}\;\dot{\cal Z}^{a}\right) dw \nonumber \\
 & = & \pd{({\cal J}_{\rm gy}\,F)}{t} + \nabla\bdot\left({\cal J}_{\rm gy}\,F\frac{}{}\dot{\bf X}\right) + \pd{}{p_{\|}}\left({\cal J}_{\rm gy}\,F\frac{}{}\dot{p}_{\|}\right) \nonumber \\
  &\equiv& {\cal J}_{\rm gy} \left( \pd{F}{t} + \dot{\bf X}\bdot\nabla F + \dot{p}_{\|}\;\pd{F}{p_{\|}}\right),
  \label{eq:Vlasov_gy}
\end{eqnarray}
where we have used the Liouville theorem \eqref{eq:Liouville} to obtain the last expression in order to recover the gyrokinetic Vlasov equation.

Next, the variation with respect to the electromagnetic potentials $(\delta\Phi_{1}, \delta{\bf A}_{1})$ yield the macroscopic gyrokinetic Maxwell equations
\begin{eqnarray}
\nabla\bdot\mathbb{D}_{\rm gy} & = & 4\pi\;\varrho_{\rm gy}, \label{eq:div_D} \\
\nabla\btimes\mathbb{H}_{\rm gy} & = & \frac{1}{c}\pd{\mathbb{D}_{\rm gy}}{t} +\; \frac{4\pi}{c} \;{\bf J}_{\rm gy}, \label{eq:curl_H} 
 \end{eqnarray}
 which can also be expressed as the microscopic Maxwell equations
\begin{eqnarray}
\nabla\bdot\epsilon\,{\bf E}_{1} & = & 4\pi\,\left(\varrho_{\rm gy} \;-\frac{}{} \nabla\bdot\mathbb{P}_{\rm gy}\right), \label{eq:div_E} \\
\nabla\btimes\left({\bf B}_{0} \;+\frac{}{} \epsilon\,{\bf B}_{1}\right) & = & \frac{\epsilon}{c}\pd{{\bf E}_{1}}{t} + \frac{4\pi}{c} \left({\bf J}_{\rm gy} + \pd{\mathbb{P}_{\rm gy}}{t} + c\,\nabla\times\mathbb{M}_{\rm gy}\right). \label{eq:curl_B} 
 \end{eqnarray}
These equations are complemented by Faraday's Law
 \begin{equation}
 \pd{{\bf B}_{1}}{t} \;+\; c\,\nabla\btimes{\bf E}_{1} \;=\; 0
 \label{eq:Faraday}
 \end{equation}
 and $\nabla\bdot{\bf B}_{1} = 0$. Now that the gyrokinetic Vlasov-Maxwell equations \eqref{eq:Vlasov_gy}-\eqref{eq:curl_H} have been derived from a variational principle, we use the remaining part of the gyrokinetic Lagrangian density variation \eqref{eq:Lag_gy_vp} to derive exact conservation laws. 
 
\section{\label{sec:symp_laws}Gyrokinetic Noether Equation and Conservation Laws}

The variational derivation of the reduced Vlasov-Maxwell equations guarantees that these reduced equations satisfy exact energy-momentum conservation laws \citep{Pfirsch_Morrison_1985,Similon_1985,CRP_2004,Brizard_2008}. In particular, the exact conservation of the gyrokinetic Vlasov-Maxwell energy \citep{Brizard_1989a,Brizard_2010a} has played an important role in the numerical implementation of the energy-conserving gyrokinetic equations \citep{Garbet_2010}. 

For this purpose, the remaining terms in Eq.~\eqref{eq:Lag_gy_vp} are combined to yield the gyrokinetic Noether equation
\begin{eqnarray}
\delta{\cal L}_{\rm gy} &=& \pd{}{t}\left(\int_{P}{\cal F}_{\rm gy}\,\delta{\cal S} - \frac{\epsilon}{4\pi c}\,\delta {\bf A}_{1}\bdot\mathbb{D}_{\rm gy}\right) \nonumber \\
 &&+\; \nabla\vb{\cdot}\left[\int_{P}\dot{\bf X}{\cal F}_{\rm gy}\delta{\cal S} - \frac{\epsilon}{4\pi}\left(\delta\Phi_{1}\mathbb{D}_{\rm gy} \;+\frac{}{} \delta {\bf A}_{1}\times\mathbb{H}_{\rm gy}\right) \right],
\label{eq:Noether_1} 
\end{eqnarray}
where the variations are now explicitly expressed in terms of the space-time displacements $\delta{\bf x}$ and $\delta t$:
\begin{equation}
\left. \begin{array}{rcl}
\delta{\cal S} & \equiv & {\bf P}_{\rm gy}\bdot\delta{\bf x} \;-\; w\,\delta t \\
\delta\Phi_{1} & \equiv & -\,\delta{\bf x}\bdot\nabla\Phi_{1} - \delta t\,\partial\Phi_{1}/\partial t \;=\; {\bf E}_{1}\bdot\delta{\bf x} \;-\; c^{-1}\partial\delta\chi_{1}/\partial t \\
\delta{\bf A}_{1} & \equiv & -\,\delta{\bf x}\bdot\nabla{\bf A}_{1} - \delta t\,\partial{\bf A}_{1}/\partial t \;=\;  {\bf E}_{1}\,c\,\delta t \;+\; \delta{\bf x}\btimes{\bf B}_{1} \;+\; \nabla\delta\chi_{1} 
\end{array} \right\},
\label{eq:delta_SphiA}
\end{equation}
with the gauge variation $\delta\chi_{1}$ defined as $\delta\chi_{1} \equiv \Phi_{1}\,c\,\delta t \;-\; {\bf A}_{1}\bdot\delta{\bf x}$. Upon rearranging the gauge variation $\delta\chi_{1}$, and using the identity
\begin{eqnarray*} 
 &&-\,\pd{}{t}\left(\nabla\delta\chi_{1}\bdot\frac{}{}\mathbb{D}_{\rm gy}\right) + \nabla\bdot\left(\pd{\delta\chi_{1}}{t}\;\mathbb{D}_{\rm gy} - c\,\nabla\delta\chi_{1}\btimes\mathbb{H}_{\rm gy}\right)  \\
 &&\;\;=\;  \pd{}{t}\left(\delta\chi_{1}\frac{}{}\nabla\bdot\mathbb{D}_{\rm gy}\right) \;-\; \nabla\bdot\left[ \delta\chi_{1}\left( \pd{\mathbb{D}_{\rm gy}}{t} - c\,\nabla\btimes\mathbb{H}_{\rm gy}\right) \right],
 \end{eqnarray*}
with the macroscopic gyrokinetic Maxwell equations \eqref{eq:div_D}-\eqref{eq:curl_H}, we obtain the gauge-invariant form of the gyrokinetic Noether equation \eqref{eq:Noether_1}:
\begin{equation} 
\delta{\cal L}_{\rm gy} \;=\; \partial\delta{\cal N}_{\rm gy}/\partial t + \nabla\bdot\delta\vb{\Gamma}_{\rm gy},
\label{eq:Noether_gy}
\end{equation}
where the action-density variation is
\begin{equation}
\delta{\cal N}_{\rm gy} \;=\; \int_{P}{\cal F}_{\rm gy}\,\left( \delta{\cal S} + \epsilon\,\frac{e}{c}\delta\chi_{1{\rm gy}}\right) \;-\; \left(\epsilon\,{\bf E}_{1}\,\delta t + \delta{\bf x}\btimes\frac{\epsilon}{c}\,{\bf B}_{1}\right) \bdot\frac{\mathbb{D}_{\rm gy}}{4\pi} ,
 \end{equation}
and the action-density-flux variation is
\begin{eqnarray}
\delta\vb{\Gamma}_{\rm gy} &=& \int_{P}\dot{\bf X}\;{\cal F}_{\rm gy}\left( \delta{\cal S} + \epsilon\,\frac{e}{c}\delta\chi_{1{\rm gy}}\right) \;-\; \delta{\bf x}\bdot\left( \frac{\epsilon}{4\pi}{\bf E}_{1}\mathbb{D}_{\rm gy} \right) \nonumber \\
 &&+\; \frac{\epsilon}{4\pi}\left({\bf E}_{1}\,c\,\delta t \;+\frac{}{} \delta{\bf x}\btimes{\bf B}_{1}\right)\times\mathbb{H}_{\rm gy}.
\end{eqnarray}
Here, the gauge-invariant terms are
\begin{eqnarray}
\delta{\cal S} + \epsilon\,\frac{e}{c}\delta\chi_{1{\rm gy}} &=& \left( {\bf P}_{\rm gy} - \epsilon\frac{e}{c}\,{\bf A}_{1{\rm gy}}\right)\bdot\delta{\bf x} \;-\; \left( w \;-\frac{}{} \epsilon\,e\,\Phi_{1{\rm gy}}\right)\,\delta t \nonumber \\
 &=& \left( \frac{e}{c}\,{\bf A}_{0}^{*}  \;+\; \vb{\Pi}_{\rm gy}\right)\bdot\delta{\bf x} \;-\; \left(K_{\rm gy} \;-\; {\cal H}_{\rm gy}\right)\,\delta t.
\end{eqnarray}
We note that the guiding-center vector potential ${\bf A}_{0}^{*}$, which yields the unperturbed equilibrium magnetic field ${\bf B}_{0}^{*} = \nabla\btimes{\bf A}_{0}^{*}$, is not subject to a gauge transformation.
 
A complete expression for the gyrokinetic Noether equation \eqref{eq:Noether_gy} also requires an explicit expression for the Lagrangian variation $\delta{\cal L}_{\rm gy}$ on the left side of Eq.~\eqref{eq:Noether_gy}. For the derivation of the momentum-energy conservation laws, we consider the specific space-time variations of the gyrokinetic Lagrangian density
\begin{eqnarray}
\delta{\cal L}_{\rm gy} &=& -\,\left(\delta t \pd{}{t} + \delta{\bf x}\bdot\nabla\right)\left[ \frac{1}{8\pi} \left( \epsilon^{2}|{\bf E}_{1}|^{2} \;-\frac{}{} |{\bf B}|^{2}\right) \right]  \nonumber \\
 &&+\; \delta{\bf x}\bdot\left[ \int_{\bf P} {\cal J}_{\rm gy}\,F \left(\nabla^{\prime}{\bf P}_{\rm gy}\bdot\dot{\bf X} \;-\; \nabla^{\prime}K_{\rm gy}\right) \;-\; \nabla{\bf B}_{0}\bdot\frac{\bf B}{4\pi}  \right],
 \label{eq:delta_Lgy_Noether}
\end{eqnarray}
where the gradient operator $\nabla^{\prime}$ only takes into account the non-uniformity of the equilibrium magnetic field, i.e., the first-order fields $(\Phi_{1},{\bf A}_{1},{\bf E}_{1},{\bf B}_{1})$ are frozen at a fixed position 
${\bf x}$ so that, for example, $\nabla^{\prime}\Phi_{1{\rm gy}} = 0$ and $\nabla^{\prime}\langle\langle B_{1\|{\rm gc}}\rangle\rangle = \nabla\bhat_{0}\bdot\langle\langle{\bf B}_{1{\rm gc}}\rangle\rangle$. It is in the second line of Eq.~\eqref{eq:delta_Lgy_Noether} that the Noether Theorem draws its full force. In addition, the $w$-integration was performed to leave the standard gyrocenter Vlasov distribution $F({\bf X},p_{\|},\mu,t)$, with $\int_{\bf P}$ now denoting an integration over $(p_{\|},\mu)$.

The final form of the gyrokinetic Noether equation is obtained by equating Eqs.~\eqref{eq:Noether_gy} and \eqref{eq:delta_Lgy_Noether}, where the virtual space-time displacements $(\delta{\bf x}, \delta t)$ appear explicitly: 
\begin{eqnarray}
&&\pd{}{t}\left[ \delta{\cal N}_{\rm gy} + \frac{\delta t}{8\pi} \left( \epsilon^{2}|{\bf E}_{1}|^{2} \;-\frac{}{} |{\bf B}|^{2}\right) \right] \;+\; \nabla\bdot\left[\delta\vb{\Gamma}_{\rm gy} + \frac{\delta{\bf x}}{8\pi} \left( \epsilon^{2}|{\bf E}_{1}|^{2} \;-\frac{}{} 
|{\bf B}|^{2}\right) \right] \nonumber \\
 &&=\; \delta{\bf x}\bdot\left[ \int_{\bf P} {\cal J}_{\rm gy}\,F \left(\nabla^{\prime}{\bf P}_{\rm gy}\bdot\dot{\bf X} \;-\; \nabla^{\prime}K_{\rm gy}\right) \;-\; \nabla{\bf B}_{0}\bdot\frac{\bf B}{4\pi}  \right].
 \label{eq:gyrokinetic_Noether}
\end{eqnarray}
This form of the Noether Theorem relies on the constrained variations \eqref{eq:delta_EB}, \eqref{eq:delta_Fgy}, and \eqref{eq:delta_SphiA}, which is in contrast to the more traditional formulation based on the connection between conservation laws and symmetries of the Vlasov-Maxwell Lagrangian (see, for example, \cite{EH_2020} and references therein). Here, for each conservation law derived from our gyrokinetic Noether equation \eqref{eq:gyrokinetic_Noether}, we also present an explicit proof based on the gyrokinetic Vlasov-Maxwell equations \eqref{eq:Vlasov_gy}-\eqref{eq:curl_H}.

\subsection{Gyrokinetic energy conservation law}

Since the equilibrium magnetic field ${\bf B}_{0}$ is time-independent, the total energy associated with the gyrokinetic Vlasov-Maxwell equations \eqref{eq:Vlasov_gy}-\eqref{eq:curl_H} is conserved. We derive the energy conservation law from the gyrokinetic Noether equation by setting $\delta t \neq 0$ and $\delta{\bf x} = 0$ in Eq.~\eqref{eq:gyrokinetic_Noether}, which yields the gyrokinetic energy conservation law 
\begin{equation}
\partial{\cal E}_{\rm gy}/\partial t \;+\; \nabla\bdot{\bf S}_{\rm gy} \;=\; 0, 
\label{eq:energy_gy}
\end{equation}
where the gyrokinetic energy density is
\begin{eqnarray}
{\cal E}_{\rm gy} & = & \int_{\bf P} {\cal J}_{\rm gy}\,F\;K_{\rm gy} + \frac{\epsilon}{4\pi}\,{\bf E}_{1}\bdot\mathbb{D}_{\rm gy} - \frac{1}{8\pi}\,\left( \epsilon^{2}|{\bf E}_{1}|^{2} \;-\frac{}{} |{\bf B}|^{2} \right) \nonumber \\
 & = &  \int_{\bf P} {\cal J}_{\rm gy}\,F\left[ \frac{p_{\|}^{2}}{2m} \;+\; \mu\,\left( B_{0} \;+\frac{}{} \epsilon\,\langle\langle B_{1\|{\rm gc}}\rangle\rangle \;+\; \frac{\epsilon^{2}}{2}\,\frac{|{\bf B}_{1}|^{2}}{B_{0}} \right) \right. \nonumber \\
  &&\left.+\; {\bf E}_{1}({\bf x})\bdot\left(\fd{\vb{\Pi}_{\rm gy}}{{\bf E}_{1}({\bf x})} \;-\; \fd{K_{\rm gy}}{{\bf E}_{1}({\bf x})}\right)\right]  \;+\; \frac{1}{8\pi}\,\left( \epsilon^{2}|{\bf E}_{1}|^{2} \;+\frac{}{} |{\bf B}|^{2} \right),
 \label{eq:E_density}
\end{eqnarray}
while the gyrokinetic energy-density flux is
\begin{equation}
{\bf S}_{\rm gy} \;=\; \int_{\bf P} {\cal J}_{\rm gy}\,F\;\dot{\bf X}\,K_{\rm gy} + \frac{c}{4\pi}\,\epsilon\,{\bf E}_{1}\btimes\mathbb{H}_{\rm gy},
 \label{eq:S_density}
 \end{equation}
 where the polarization and magnetization $(\mathbb{P}_{\rm gy}, \mathbb{M}_{\rm gy})$ are defined in Eqs.~\eqref{eq:Pol_gy}-\eqref{eq:Mag_gy}, with $\mathbb{H}_{\rm gy}$ defined in Eq.~\eqref{eq:DH_def}. In addition, we note that the gyrokinetic polarization  
and magnetization $(\mathbb{P}_{\rm gy}, \mathbb{M}_{\rm gy})$ include the full gyrocenter velocity $\dot{\bf X}$ defined in Eq.~\eqref{eq:Xgy_dot}, which is expressed in terms of the effective electric and magnetic fields \eqref{eq:Egy_*}-\eqref{eq:Bgy_*}. We also note that, as shown by \cite{Burby_2015}, the gyrokinetic Vlasov-Maxwell Hamiltonian functional is naturally derived from the gyrokinetic energy density \eqref{eq:E_density}.
 
 The explicit proof of gyrokinetic energy conservation, which applies to both gauge-free gyrokinetic models \citep{Burby_Brizard_2019,Brizard_2020} considered here, proceeds as follows. First, we begin with the partial time derivative of the gyrokinetic energy density \eqref{eq:E_density}:
 \begin{eqnarray}
 \pd{{\cal E}_{\rm gy}}{t} &=& \int_{\bf P} \left[ \pd{({\cal J}_{\rm gy}\,F)}{t}\;K_{\rm gy} \;+\; {\cal J}_{\rm gy}\,F\;\left( \pd{{\bf E}_{1}}{t}\bdot\pd{K_{\rm gy}}{{\bf E}_{1}} + \pd{{\bf B}_{1}}{t}\bdot\pd{K_{\rm gy}}{{\bf B}_{1}} \right) \right] \nonumber \\
  &&+\; \frac{\epsilon\,{\bf E}_{1}}{4\pi}\bdot\pd{\mathbb{D}_{\rm gy}}{t} \;+\; \epsilon\,\pd{{\bf E}_{1}}{t}\bdot\mathbb{P}_{\rm gy} \;+\; \frac{\bf B}{4\pi}\bdot\epsilon\,\pd{{\bf B}_{1}}{t},
 \label{eq:Eproof_1}
  \end{eqnarray}
where we expanded the term $\partial K_{\rm gy}({\bf E}_{1},{\bf B}_{1})/\partial t$ and used the definition \eqref{eq:DH_def} for $\mathbb{D}_{\rm gy}$. Using the phase-space divergence form \eqref{eq:Vlasov_gy} of the gyrokinetic Vlasov equation, the first term on the right can be expressed as
 \begin{equation}
 \int_{\bf P} \pd{({\cal J}_{\rm gy}\,F)}{t}\;K_{\rm gy} \;=\; -\;\nabla\bdot\left(\int_{\bf P}{\cal J}_{\rm gy}\,F\,K_{\rm gy}\;\dot{\bf X}\right) \;+\; \int_{\bf P}{\cal J}_{\rm gy}\,F \left( \pd{K_{\rm gy}}{p_{\|}}\;\dot{p}_{\|} \;+\; \dot{\bf X}\bdot\nabla K_{\rm gy}\right),
  \end{equation}
 while, using the definitions \eqref{eq:Pol_gy}-\eqref{eq:Mag_gy} of the gyrokinetic polarization and magnetization, the gyrokinetic kinetic terms in Eq.~\eqref{eq:Eproof_1} can be expressed
\[ \int_{\bf P} {\cal J}_{\rm gy}\,F\;\pd{K_{\rm gy}}{t} \;=\; -\,\epsilon\,\left(\pd{{\bf E}_{1}}{t}\vb{\cdot}\mathbb{P}_{\rm gy}  + \pd{{\bf B}_{1}}{t}\vb{\cdot}\mathbb{M}_{\rm gy} \right) + \int_{\bf P} {\cal J}_{\rm gy}\,F\;\left(
\pd{\vb{\Pi}_{\rm gy}}{t}\vb{\cdot}\dot{\bf X}\right) \]
By combining these expressions, Eq.~\eqref{eq:Eproof_1} becomes
\begin{eqnarray}
 \pd{{\cal E}_{\rm gy}}{t} & = & -\;\nabla\bdot\left(\int_{\bf P}{\cal J}_{\rm gy}\,F\,K_{\rm gy}\;\dot{\bf X}\right) + \frac{\mathbb{H}_{\rm gy}}{4\pi}\bdot\epsilon\,\pd{{\bf B}_{1}}{t} + \frac{\epsilon\,{\bf E}_{1}}{4\pi}\bdot\pd{\mathbb{D}_{\rm gy}}{t} 
   \label{eq:Eproof_2} \\
  & &+ \int_{\bf P}{\cal J}_{\rm gy}\,F\left[\pd{K_{\rm gy}}{p_{\|}}\;\dot{p}_{\|} \;+\;\dot{\bf X}\bdot\left( \nabla K_{\rm gy} \;+\; \pd{\vb{\Pi}_{\rm gy}}{t}\right) \right],
\nonumber
\end{eqnarray}
where we introduced the definition \eqref{eq:DH_def} for $\mathbb{H}_{\rm gy}$. Next, we use Faraday's Law \eqref{eq:Faraday} to write
\[ \frac{\mathbb{H}_{\rm gy}}{4\pi}\bdot\epsilon\,\pd{{\bf B}_{1}}{t}  \;=\; -\;\frac{c\,\mathbb{H}_{\rm gy}}{4\pi}\bdot\nabla\btimes\epsilon\,{\bf E}_{1} \;=\; -\;\nabla\bdot\left( \frac{c}{4\pi}\,\epsilon\,{\bf E}_{1}\btimes\mathbb{H}_{\rm gy}\right) \;-\; 
\frac{\epsilon\,{\bf E}_{1}}{4\pi}\bdot c\,\nabla\btimes\mathbb{H}_{\rm gy}, \]
so that Eq.~\eqref{eq:Eproof_2} becomes
\begin{eqnarray}
\pd{{\cal E}_{\rm gy}}{t} + \nabla\bdot{\bf S}_{\rm gy} &=& -\;\frac{\epsilon\,{\bf E}_{1}}{4\pi}\bdot\left( c\,\nabla\btimes\mathbb{H}_{\rm gy} \;-\; \pd{\mathbb{D}_{\rm gy}}{t} \right) \nonumber \\
 &&+\; \int_{\bf P}{\cal J}_{\rm gy}\,F\left[ \pd{K_{\rm gy}}{p_{\|}}\;\dot{p}_{\|} \;+\;   \dot{\bf X}\bdot\left( \nabla K_{\rm gy} \;+\; \pd{\vb{\Pi}_{\rm gy}}{t}\right) \right],
 \label{eq:Eproof_3}
 \end{eqnarray}
where we reconstructed the gyrokinetic energy-density flux \eqref{eq:S_density} on the left side of Eq.~\eqref{eq:Eproof_3}. Lastly, we use the identity derived from Eq.~\eqref{eq:Egy_*}:
\begin{equation}
\nabla K_{\rm gy} \;+\; \pd{\vb{\Pi}_{\rm gy}}{t} \;=\; e\,\left( \epsilon\;{\bf E}_{1{\rm gy}} \;-\; {\bf E}_{\rm gy}^{*}\right), 
\label{eq:K_Pi_id}
\end{equation}
and we use the macroscopic gyrokinetic Maxwell equation \eqref{eq:curl_H}, with
\[ -\;\frac{\epsilon\,{\bf E}_{1}}{4\pi}\bdot\left( c\,\nabla\btimes\mathbb{H}_{\rm gy} \;-\; \pd{\mathbb{D}_{\rm gy}}{t} \right) \;=\; -\;\epsilon\,{\bf E}_{1}\bdot{\bf J}_{\rm gy} \;=\; -\;\int_{\bf P}{\cal J}_{\rm gy}\,F\left(\epsilon\,e\,{\bf E}_{1{\rm gy}}\bdot\dot{\bf X}
\right), \]
 to obtain
\begin{equation}
\pd{{\cal E}_{\rm gy}}{t} + \nabla\bdot{\bf S}_{\rm gy} \;=\;  \int_{\bf P}{\cal J}_{\rm gy}\,F\left( \pd{K_{\rm gy}}{p_{\|}}\;\dot{p}_{\|} \;-\; \dot{\bf X}\bdot\,e\,{\bf E}_{\rm gy}^{*} \right).
\label{eq:E_proof4}
 \end{equation}
Using the Euler-Lagrange identity \eqref{eq:xp_dot_id}, the right side of Eq.~\eqref{eq:E_proof4} is shown to vanish identically and we readily recover the exact gyrokinetic energy conservation law. 
 
\subsection{Gyrokinetic Noether momentum equation}

Because the equilibrium magnetic field ${\bf B}_{0}$ considered in standard gyrokinetic Vlasov-Maxwell theory is spatially non-uniform (i.e., it serves to magnetically confine charged particles in accordance with the guiding-center approximation), a general gyrokinetic Vlasov-Maxwell momentum conservation law does not exist. Indeed, according to the Noether Theorem, the gyrokinetic Vlasov-Maxwell momentum is conserved only in directions corresponding to symmetries of the equilibrium magnetic field. Before we derive the gyrokinetic angular-momentum conservation law associated with an axisymmetric equilibrium magnetic field, we wish to show that the gyrokinetic Noether momentum equation, from which our exact angular-momentum conservation law will be derived, is consistent with the gyrokinetic Vlasov-Maxwell equations \eqref{eq:Vlasov_gy}-\eqref{eq:curl_H}.

We begin with the gyrokinetic Noether momentum equation derived by setting $\delta t = 0$ and $\delta{\bf x} \neq 0$ in Eq.~\eqref{eq:gyrokinetic_Noether}:
\begin{equation}
 \pd{\vb{\cal P}^{*}_{\rm gy}}{t} + \nabla\bdot{\sf T}^{*}_{\rm gy} =  \int_{\bf P} {\cal J}_{\rm gy}\,F\left( \frac{e}{c}\nabla{\bf A}_{0}^{*}\bdot\dot{\bf X} + \nabla^{\prime}\vb{\Pi}_{\rm gy}\bdot\dot{\bf X} - \nabla^{\prime}K_{\rm gy}\right)  - 
 \nabla{\bf B}_{0}\bdot\frac{\bf B}{4\pi},
 \label{eq:momentum}
 \end{equation}
where the gyrokinetic  canonical momentum density is defined as
\begin{equation} 
\vb{\cal P}^{*}_{\rm gy} \;=\; \int_{\bf P} {\cal J}_{\rm gy}\,F\,\left( \frac{e}{c}\,{\bf A}_{0}^{*} \;+\; \vb{\Pi}_{\rm gy}\right) \;+\; \frac{\mathbb{D}_{\rm gy}}{4\pi c}\btimes\epsilon\,{\bf B}_{1} 
\label{eq:Pgy_can}
\end{equation}
and the gyrokinetic  canonical stress tensor is defined as
\begin{eqnarray}
{\sf T}^{*}_{\rm gy} & = &  \int_{\bf P} {\cal J}_{\rm gy}\,F\;\dot{\bf X}\,\left( \frac{e}{c}\,{\bf A}_{0}^{*} \;+\;\vb{\Pi}_{\rm gy}\right) \;-\; \frac{\epsilon}{4\pi} \left( \mathbb{D}_{\rm gy}\,{\bf E}_{1} \;+\frac{}{} {\bf B}_{1}\,\mathbb{H}_{\rm gy}\right) \nonumber \\
 &  &+\; {\bf I} \left[ \frac{1}{8\pi} \left( \epsilon^{2}\,|{\bf E}_{1}|^{2} \;-\frac{}{} |{\bf B}|^{2} \right) + \frac{\epsilon}{4\pi}\,{\bf B}_{1}\bdot\mathbb{H}_{\rm gy}\right],
\label{eq:Tgy_can}
\end{eqnarray}
where $ {\bf I}$ denotes the identity matrix. We note that, while the gyrokinetic stress tensor \eqref{eq:Tgy_can} is manifestly not symmetric, the conservation of the gyrokinetic angular-momentum will follow exactly from Eq.~\eqref{eq:momentum}.

 We would now like to show that Eq.~\eqref{eq:momentum} is an exact consequence of the gyrokinetic Vlasov-Maxwell equations \eqref{eq:Vlasov_gy}-\eqref{eq:curl_H}. We begin with the partial time derivatives of the first two terms in the gyrokinetic canonical momentum density \eqref{eq:Pgy_can}:
\begin{eqnarray*} 
\pd{}{t}\left( \int_{\bf P} {\cal J}_{\rm gy}\,F\;\frac{e}{c}\,{\bf A}_{0}^{*}\right) &=& -\;\nabla\bdot \left(\int_{\bf P} {\cal J}_{\rm gy}\,F\;\dot{\bf X}\;\frac{e}{c}\,{\bf A}_{0}^{*}\right) \\
 &&+  \int_{\bf P} {\cal J}_{\rm gy}\,F\; \left( \frac{e}{c}\nabla{\bf A}_{0}^{*}\bdot\dot{\bf X} + \dot{p}_{\|}\;\bhat_{0} - \frac{e}{c}\dot{\bf X}\btimes{\bf B}_{0}^{*}\right), \\
\pd{}{t}\left( \int_{\bf P} {\cal J}_{\rm gy}\,F\;\vb{\Pi}_{\rm gy}\right) &=& -\;\nabla\bdot\left(\int_{\bf P} {\cal J}_{\rm gy}\,F\;\dot{\bf X}\;\vb{\Pi}_{\rm gy} \right) \;+\; \int_{\bf P}{\cal J}_{\rm gy}F \left( \nabla\vb{\Pi}_{\rm gy}\bdot\dot{\bf X}  \;-\frac{}{} 
\nabla K_{\rm gy} \right) \\
 &&+\; \int_{\bf P}{\cal J}_{\rm gy}F \left[ \dot{p}_{\|}\;\pd{\vb{\Pi}_{\rm gy}}{p_{\|}} + e \left(\epsilon\,{\bf E}_{1{\rm gy}} - {\bf E}_{\rm gy}^{*}\right) \;-\; \dot{\bf X}\btimes\nabla\btimes\vb{\Pi}_{\rm gy}
 \right],
\end{eqnarray*}
where we used the phase-space divergence form \eqref{eq:Vlasov_gy} of the gyrokinetic Vlasov equation, followed by integrations by parts, and used Eq.~\eqref{eq:K_Pi_id} to write $\partial\vb{\Pi}_{\rm gy}/\partial t$. By combining these two expressions, we obtain
\begin{eqnarray}
\pd{}{t}\left[  \int_{\bf P}{\cal J}_{\rm gy}F \left(\frac{e}{c}\,{\bf A}_{0}^{*} \;+\; \vb{\Pi}_{\rm gy}\right)\right] & = & -\;\nabla\bdot\left[  \int_{\bf P}{\cal J}_{\rm gy}F\,\dot{\bf X} \left(\frac{e}{c}\,{\bf A}_{0}^{*} \;+\; \vb{\Pi}_{\rm gy}\right)\right] \label{eq:P_proof_1} \\
 &&+\;  \int_{\bf P}{\cal J}_{\rm gy}F \left[ \left(\frac{e}{c}\,\nabla{\bf A}_{0}^{*} \;+\; \nabla\vb{\Pi}_{\rm gy}\right)\bdot\dot{\bf X} \;-\; \nabla K_{\rm gy} \right] \nonumber \\
 &&+\;  \int_{\bf P}{\cal J}_{\rm gy}F \left[ \epsilon \left( e\,{\bf E}_{1{\rm gy}} \;+\; \frac{e}{c}\,\dot{\bf X}\btimes{\bf B}_{1{\rm gy}} \right) \right] \nonumber \\
 &&+\; \int_{\bf P}{\cal J}_{\rm gy}F \left[ \dot{p}_{\|}\;{\sf b}_{\rm gy}^{*} \;-\; \left( e\,{\bf E}_{\rm gy}^{*} \;+\; \frac{e}{c}\,\dot{\bf X}\btimes{\bf B}_{\rm gy}^{*} \right) \right], \nonumber 
\end{eqnarray}
where the last line vanishes as a result of the gyrocenter Euler-Lagrange equation \eqref{eq:EL_X}. Next, we take the partial time derivative of the third term in Eq.~\eqref{eq:Pgy_can}:
\begin{eqnarray}
\pd{}{t}\left( \frac{\mathbb{D}_{\rm gy}}{4\pi c}\btimes\epsilon\,{\bf B}_{1} \right) & = & \frac{1}{4\pi}\left( \frac{1}{c}\pd{\mathbb{D}_{\rm gy}}{t}\btimes\epsilon\,{\bf B}_{1} + \mathbb{D}_{\rm gy}\btimes\frac{\epsilon}{c}\pd{{\bf B}_{1}}{t}\right) \nonumber \\
 &=& \left(\nabla\btimes\mathbb{H}_{\rm gy}\right)\btimes\frac{\epsilon\,{\bf B}_{1}}{4\pi} - \frac{\mathbb{D}_{\rm gy}}{4\pi}\btimes\left(\nabla\btimes\epsilon\,{\bf E}_{1} \right) - \int_{\bf P}{\cal J}_{\rm gy}F \left( \frac{e}{c}\dot{\bf X}\btimes 
 \epsilon\,{\bf B}_{1{\rm gy}}\right) \nonumber \\
  &=& \nabla\bdot\left[ \frac{\epsilon}{4\pi}\,\left({\bf B}_{1}\,\mathbb{H}_{\rm gy} + \mathbb{D}_{\rm gy}\,{\bf E}_{1}\right) - \frac{\bf I}{4\pi} \left( \epsilon{\bf B}_{1}\bdot\mathbb{H}_{\rm gy} + \frac{\epsilon^{2}}{2}\,|{\bf E}_{1}|^{2} - \frac{1}{2} 
  |{\bf B}|^{2} \right) \right] \nonumber \\
 &&-\; \epsilon\left(\nabla{\bf E}_{1}\bdot\mathbb{P}_{\rm gy} \;+\frac{}{} \nabla{\bf B}_{1}\bdot\mathbb{M}_{\rm gy}\right) \;-\; \nabla{\bf B}_{0}\bdot\frac{\bf B}{4\pi} \nonumber \\
  &&-\; \int_{\bf P}{\cal J}_{\rm gy}F \,\epsilon\,\left( e\,{\bf E}_{1{\rm gy}} \;+\; \frac{e}{c}\dot{\bf X}\btimes {\bf B}_{1{\rm gy}}\right).
  \label{eq:P_proof_2}
\end{eqnarray}
When we combine Eqs.~\eqref{eq:P_proof_1}-\eqref{eq:P_proof_2}, we obtain
\begin{eqnarray}
\pd{\vb{\cal P}^{*}_{\rm gy}}{t} + \nabla\bdot{\sf T}^{*}_{\rm gy} &=& \int_{\bf P}{\cal J}_{\rm gy}F \left[ \left(\frac{e}{c}\,\nabla{\bf A}_{0}^{*} \;+\; \nabla\vb{\Pi}_{\rm gy}\right)\bdot\dot{\bf X} \;-\; \nabla K_{\rm gy} \right] \nonumber \\
 &&-\; \epsilon\left(\nabla{\bf E}_{1}\bdot\mathbb{P}_{\rm gy} \;+\frac{}{} \nabla{\bf B}_{1}\bdot\mathbb{M}_{\rm gy}\right) \;-\; \nabla{\bf B}_{0}\bdot\frac{\bf B}{4\pi},
 \label{eq:P_proof_3}
\end{eqnarray}
where
\begin{eqnarray*} 
-\,\epsilon\left(\nabla{\bf E}_{1}\bdot\mathbb{P}_{\rm gy} \;+\frac{}{} \nabla{\bf B}_{1}\bdot\mathbb{M}_{\rm gy}\right) &=& -\,\int_{\bf P}{\cal J}_{\rm gy}F \left(\nabla{\bf E}_{1}\bdot\fd{\vb{\Pi}_{\rm gy}}{{\bf E}_{1}} + \nabla{\bf B}_{1}\bdot
\fd{\vb{\Pi}_{\rm gy}}{{\bf B}_{1}} \right)\bdot\dot{\bf X} \\
 &&+\;\int_{\bf P}{\cal J}_{\rm gy}F \left(\nabla{\bf E}_{1}\bdot\fd{K_{\rm gy}}{{\bf E}_{1}} + \nabla{\bf B}_{1}\bdot\fd{K_{\rm gy}}{{\bf B}_{1}} \right) \\
  &\equiv& -\,\int_{\bf P}{\cal J}_{\rm gy}F \left[ \left(\nabla\vb{\Pi}_{\rm gy} - \nabla^{\prime}\vb{\Pi}_{\rm gy}\right)\bdot\dot{\bf X} -  \left(\nabla K_{\rm gy} - \nabla^{\prime}K_{\rm gy}\right)\right].
\end{eqnarray*}
By inserting these terms in Eq.~\eqref{eq:P_proof_3}, we recover the gyrokinetic Noether momentum equation \eqref{eq:momentum}.

We note that, while the gyrokinetic Noether momentum equation \eqref{eq:momentum} is not a gyrokinetic conservation law, it can be used directly to obtain a gyrokinetic momentum transport equation (e.g., in the parallel direction) by taking its projection in the desired direction. For example, the gyrokinetic canonical parallel-momentum transport equation is expressed as
\begin{eqnarray}
\pd{{\cal P}^{*}_{{\rm gy}\|}}{t} + \nabla\bdot\left({\sf T}^{*}_{\rm gy}\bdot\bhat_{0}\right) &=& {\sf T}^{*\top}_{\rm gy}:\nabla\bhat_{0} \;-\; \bhat_{0}\bdot\nabla{\bf B}_{0}\bdot{\bf B}/4\pi 
\label{eq:momenum_par*} \\
 &&\;+\; \int_{\bf P} {\cal J}_{\rm gy}\,F\,\bhat_{0}\bdot\left( \frac{e}{c}\nabla{\bf A}_{0}^{*}\bdot\dot{\bf X} + \nabla^{\prime}\vb{\Pi}_{\rm gy}\bdot\dot{\bf X} - \nabla^{\prime}K_{\rm gy}\right), \nonumber 
 \end{eqnarray}
 where ${\cal P}^{*}_{{\rm gy}\|} \equiv \vb{\cal P}^{*}_{\rm gy}\bdot\bhat_{0}$ and ${\sf T}_{{\rm gy}}^{*\top}$ denotes the transpose of the gyrokinetic stress tensor \eqref{eq:Tgy_can}. The gyrokinetic canonical parallel-momentum transport equation
 \eqref{eq:momenum_par*} can be transformed into a simpler form as the $p_{\|}$-moment of the gyrokinetic Vlasov equation
\begin{equation}
 \pd{}{t}\left(\int_{\bf P}{\cal J}_{\rm gy}F\,p_{\|} \right) \;+\; \nabla\bdot\left(\int_{\bf P}{\cal J}_{\rm gy}F\;\dot{\bf X}\,p_{\|} \right) \;=\; \int_{\bf P}{\cal J}_{\rm gy}F\,\dot{p}_{\|},
 \label{eq:momenum_par}
 \end{equation}
 where the gyrocenter parallel force $\dot{p}_{\|}$ is defined by Eq.~\eqref{eq:pgy_dot}. See \cite{Brizard_Tronko_2011} for the explicit transformation from Eq.~\eqref{eq:momenum_par*} to Eq.~\eqref{eq:momenum_par} for the case of the gyrokinetic Vlasov-Poisson equations. We note that the parallel contraction of the gyrokinetic stress tensor ${\sf T}^{*}_{\rm gy}\bdot\bhat_{0}$ on the left side of Eq.~\eqref{eq:momenum_par*} contains the gyrokinetic Maxwell stress-tensor term $-\,\mathbb{D}_{\rm gy}\,\epsilon\,E_{1\|}/4\pi$, which plays a central role in the electrostatic gyrokinetic Vlasov-Poisson model of \cite{McDevitt_2009} in discussing toroidal rotation driven by the gyrocenter polarization $\mathbb{P}_{\rm gy}$. In particular, \cite{McDevitt_2009} show how this polarization contribution can be retrieved from a perturbation expansion (up to fourth order) of the right side of Eq.~\eqref{eq:momenum_par} through a $\delta F$-decomposition of the gyrocenter Vlasov distribution. Our gyrokinetic canonical parallel-momentum transport equation \eqref{eq:momenum_par*}, in contrast, explicitly exhibits the complete gyrocenter polarization and magnetization effects in a full-F gyrokinetic Vlasov-Maxwell theory.

\subsection{Gyrokinetic angular-momentum conservation law}

Assuming now that the equilibrium magnetic field ${\bf B}_{0}$ is axisymmetric (i.e., $\partial{\bf B}_{0}/\partial\varphi = \wh{\sf z}\btimes{\bf B}_{0}$), we derive the gyrokinetic canonical angular-momentum conservation law by taking the scalar product of Eq.~\eqref{eq:momentum} with $\partial{\bf x}/\partial\varphi$ (i.e., 
$\delta{\bf x} = \delta\varphi\;\partial{\bf x}/\partial\varphi$), where the toroidal angle $\varphi$ is associated with rotations about the $z$-axis. Hence, the toroidal canonical 
angular-momentum density ${\cal P}_{{\rm gy}\varphi}^{*} \equiv  \vb{\cal P}_{{\rm gy}}^{*}\bdot\partial{\bf x}/\partial\varphi$ satisfies the Noether canonical angular-momentum equation
\begin{eqnarray}
\pd{{\cal P}_{{\rm gy}\varphi}^{*}}{t} + \nabla\vb{\cdot}\left({\sf T}_{{\rm gy}}^{*}\bdot\pd{\bf x}{\varphi}\right) &=& {\sf T}_{{\rm gy}}^{*\top}\;\vb{:}\;\nabla\left(\pd{\bf x}{\varphi}\right) - \pd{{\bf B}_{0}}{\varphi}\bdot\frac{\bf B}{4\pi} \nonumber \\
 &&+ \int_{\bf P} {\cal J}_{\rm gy}\,F\left( \frac{e}{c}\,\pd{{\bf A}_{0}^{*}}{\varphi}\bdot\dot{\bf X} + \frac{\partial^{\prime}\vb{\Pi}_{\rm gy}}{\partial\varphi}\bdot\dot{\bf X} - \frac{\partial^{\prime}K_{\rm gy}}{\partial\varphi}\right),
 \label{eq:Pphi}
 \end{eqnarray}
Under the assumption that the equilibrium magnetic field is axisymmetric, we have $\partial B_{0}/\partial\varphi \equiv 0$ and we will use the identity $\partial\bhat_{0}/\partial\varphi \equiv \wh{\sf z}\btimes\bhat_{0}$, so that ${\bf B}\bdot\partial{\bf B}_{0}/\partial\varphi = \epsilon\,{\bf B}_{1}\bdot(\wh{\sf z}\btimes{\bf B}_{0})$. 

Instead of merely assuming that the right side of Eq.~\eqref{eq:Pphi} is zero, we will now systematically show how the various terms do cancel each other out to yield an exact conservation law. Before we begin, however, we note that the first term vanishes identically if the gyrokinetic stress tensor \eqref{eq:Tgy_can} is symmetric (i.e., ${\sf T}_{{\rm gy}}^{*\top} = {\sf T}_{{\rm gy}}^{*}$), which is expected (and required) when there is no separation between dynamical fields and equilibrium fields, e.g., in guiding-center Vlasov-Maxwell theory \citep{Brizard_Tronci_2016}.  In the present case, however, the asymmetry of the gyrokinetic stress tensor \eqref{eq:Tgy_can} is necessary in order to cancel the additional terms on the right of Eq.~\eqref{eq:Pphi}.

We now proceed with the proof that the right side of Eq.~\eqref{eq:Pphi} is zero for the gauge-free model of \cite{Burby_Brizard_2019}, where $\vb{\Pi}_{\rm gy} \equiv 0$, and present the results for the gauge-free model of \cite{Brizard_2020}. First, we note that since the dyadic tensor $\nabla(\partial{\bf x}/\partial\varphi) = \wh{R}\,\wh{\varphi} - \wh{\varphi}\,\wh{R}$ is  anti-symmetric (where $R \equiv |\partial{\bf x}/\partial\varphi|$), only the anti-symmetric part of ${\sf T}_{{\rm gy}}^{*\top}$ contributes in the first term of Eq.~\eqref{eq:Pphi}:
\begin{eqnarray}
{\sf T}_{{\rm gy}}^{*\top}\;\vb{:}\;\nabla\left(\pd{\bf x}{\varphi}\right) & = & \wh{\sf z}\bdot\left[ \int_{\bf P} {\cal J}_{\rm gy}\,F\;\left(\dot{\bf X}\btimes\frac{e}{c}\,{\bf A}_{0}^{*}\right) \;-\; \frac{\epsilon}{4\pi} \left(\mathbb{D}_{\rm gy}\btimes
{\bf E}_{1} \;+\frac{}{} {\bf B}_{1}\btimes\mathbb{H}_{\rm gy}\right) \right] \nonumber \\
 & = & \wh{\sf z}\bdot\left[ \int_{\bf P} {\cal J}_{\rm gy}\,F\;\left(\dot{\bf X}\btimes\frac{e}{c}\,{\bf A}_{0}^{*}\right) \;+\; \epsilon\,{\bf E}_{1}\btimes\mathbb{P}_{\rm gy} \;+\; \epsilon\,{\bf B}_{1}\btimes\mathbb{M}_{\rm gy}\right] 
 \nonumber \\
  &&-\; \frac{\wh{\sf z}}{4\pi}\bdot(\epsilon\,{\bf B}_{1}\btimes{\bf B}_{0}),
\end{eqnarray} 
where we used the dyadic identities ${\bf I}:\nabla(\partial{\bf x}/\partial\varphi) = \nabla\bdot(\partial{\bf x}/\partial\varphi) = 0$ and ${\bf V}{\bf W}:\nabla(\partial{\bf x}/\partial\varphi) \equiv \wh{\sf z}\bdot({\bf W}\btimes{\bf V})$, which holds for an arbitrary pair of vectors $({\bf V},{\bf W})$. Next, the last term is
\begin{equation}
\frac{\partial^{\prime}K_{\rm gy}}{\partial\varphi} = \wh{\sf z}\bdot \left[\epsilon\,\mu\,\bhat_{0}\btimes\langle\langle {\bf B}_{1{\rm gc}}\rangle\rangle + \frac{p_{\|}\bhat_{0}}{mc}\btimes\left(\epsilon\,{\bf B}_{1}\btimes\vb{\pi}_{\rm gy} \right) +
\vb{\pi}_{\rm gy}\btimes\epsilon\left( {\bf E}_{1} + \frac{p_{\|}\bhat_{0}}{mc}\btimes{\bf B}_{1}\right) \right],
\end{equation}
where $\vb{\pi}_{\rm gy} \equiv \vb{\pi}_{\rm gc} + \epsilon\,\vb{\pi}_{2}$. Lastly, we write $\partial{\bf A}_{0}^{*}/\partial\varphi = \wh{\sf z}\btimes{\bf A}_{0}^{*}$ and, after some cancellations, Eq.~\eqref{eq:Pphi} becomes
\begin{equation}
 \pd{{\cal P}_{{\rm gy}\varphi}^{*}}{t} \;+\; \nabla\bdot\left({\sf T}_{{\rm gy}}^{*}\bdot\pd{\bf x}{\varphi}\right) = \wh{\sf z}\bdot\epsilon\left({\bf E}_{1}\btimes\mathbb{P}_{\rm gy} \;+\frac{}{} {\bf B}_{1}\btimes\mathbb{M}_{\rm gy}\right) \;-\;
\int_{\bf P} {\cal J}_{\rm gy}\,F\;\frac{\partial^{\prime}K_{\rm gy}}{\partial\varphi},
\label{eq:Pphi_2}
\end{equation}
where
\begin{eqnarray*}
\wh{\sf z}\bdot\epsilon\left({\bf E}_{1}\btimes\mathbb{P}_{\rm gy} \;+\frac{}{} {\bf B}_{1}\btimes\mathbb{M}_{\rm gy}\right) & = & \int_{\bf P} {\cal J}_{\rm gy}\,F\;\wh{\sf z}\bdot\epsilon\left( {\bf E}_{1}\btimes\vb{\pi}_{\rm gy} \;-\; \mu\,
\langle\langle{\bf B}_{1{\rm gc}}\rangle\rangle\btimes\bhat_{0} \right) \\
 &&+\; \int_{\bf P} {\cal J}_{\rm gy}\,F\;\wh{\sf z}\bdot\left[ {\bf B}_{1}\btimes\left(\vb{\pi}_{\rm gy}\btimes\frac{p_{\|}\bhat_{0}}{mc}\right) \right]
 \end{eqnarray*}
Upon further cancellations, Eq.~\eqref{eq:Pphi_2} becomes
\begin{equation} 
\pd{{\cal P}_{{\rm gy}\varphi}^{*}}{t} + \nabla\bdot\left({\sf T}_{{\rm gy}}^{*}\vb{\cdot}\pd{\bf x}{\varphi}\right) \;=\; \int_{\bf P} {\cal J}_{\rm gy}\,F\;\wh{\sf z}\bdot{\bf N},
\end{equation}
where the gyrocenter torque
\begin{equation}
{\bf N} \equiv \epsilon\,\frac{p_{\|}}{mc}\left[ {\bf B}_{1}\vb{\times}\left(\vb{\pi}_{\rm gy}\vb{\times}\bhat_{0}\right) + \vb{\pi}_{\rm gy}\vb{\times}\left(\bhat_{0}\vb{\times} {\bf B}_{1}\right) + \bhat_{0}\vb{\times}\left( {\bf B}_{1}\btimes\frac{}{}\vb{\pi}_{\rm gy}\right)\right] \;\equiv\; 0
\end{equation}
vanishes according to the Jacobi identity 
\begin{equation}
{\bf U}\vb{\times}({\bf V}\vb{\times}{\bf W}) \;+\; {\bf V}\vb{\times}({\bf W}\vb{\times}{\bf U}) \;+\; {\bf W}\vb{\times}({\bf U}\vb{\times}{\bf V}) \;\equiv\; 0
\label{eq:UVW_id}
\end{equation}  
for the double vector product of any three arbitrary vector fields 
$({\bf U}, {\bf V}, {\bf W})$. For the gauge-free model of \cite{Brizard_2020}, the gyrocenter torque 
\[ {\bf N} \;\equiv\; \sum_{i = 1}^{3}\left[{\bf U}_{i}\vb{\times}({\bf V}_{i}\vb{\times}{\bf W}_{i}) \;+\frac{}{} {\bf V}_{i}\vb{\times}({\bf W}_{i}\vb{\times}{\bf U}_{i}) \;+\; {\bf W}_{i}\vb{\times}({\bf U}_{i}\vb{\times}{\bf V}_{i}) \right] \;\equiv\; 0 \]
also vanishes as a result of the Jacobi vector identity \eqref{eq:UVW_id}, where 
\begin{equation}
\left. \begin{array}{rcl}
({\bf U}_{1},{\bf V}_{1},{\bf W}_{1}) &=& \left(\epsilon\langle{\bf B}_{1{\rm gc}}\rangle, \vb{\pi}_{\rm gy}, p_{\|}\bhat_{0}/mc\right) \\
({\bf U}_{2},{\bf V}_{2},{\bf W}_{2}) &=& \left(\epsilon{\bf B}_{1}, -\epsilon\,\vb{\pi}_{2}, p_{\|}\bhat_{0}/mc\right) \\
({\bf U}_{3},{\bf V}_{3},{\bf W}_{3}) &=& \left(\dot{\bf X}, \epsilon\langle{\bf E}_{1{\rm gc}}\rangle + p_{\|}\bhat_{0}/mc\btimes \epsilon\langle{\bf B}_{1{\rm gc}}\rangle,e\bhat_{0}/\Omega_{0}\right)
\end{array} \right\}.
\end{equation}

\subsection{Gyrokinetic angular-momentum conservation in axisymmetric tokamak plasmas}

Hence, we have explicitly proved that the gyrokinetic canonical angular-momentum conservation law
\begin{equation}
\pd{{\cal P}_{{\rm gy}\varphi}^{*}}{t} + \nabla\bdot\left({\sf T}_{{\rm gy}}^{*}\bdot\pd{\bf x}{\varphi}\right) \;=\; 0,
\label{eq:Pphi*_final}
\end{equation}
follows exactly from the gyrokinetic Vlasiov-Maxwell equations. We now evaluate this equation in axisymmetric tokamak geometry, in which the tokamak magnetic field is ${\bf B}_{0} = B_{0\varphi}(\psi)\,\nabla\varphi + \nabla\varphi\btimes\nabla\psi$, where $\psi$ denotes the magnetic poloidal flux and the toroidal component $B_{0\varphi}(\psi)$ is a flux function. In Eq.~\eqref{eq:Pphi*_final}, the total toroidal angular-momentum density
\begin{equation} 
{\cal P}^{*}_{{\rm gy}\varphi} \;=\; \int_{\bf P} {\cal J}_{\rm gy}\,F\,\left( P_{{\rm gc}\varphi}^{*} \;+\; \vb{\Pi}_{\rm gy}\bdot\pd{\bf x}{\varphi}  \right) \;+\; \frac{\mathbb{D}_{\rm gy}}{4\pi c}\btimes\epsilon\,{\bf B}_{1}\bdot\pd{\bf x}{\varphi}
\label{eq:Pgy_can_phi} 
\end{equation}
is the sum of three groups of terms. 

The first group in Eq.~\eqref{eq:Pgy_can_phi} is defined as the gyrocenter moment of the guiding-center toroidal angular-momentum
\begin{equation}
P_{{\rm gc}\varphi}^{*} \;\equiv\; \frac{e}{c}\,{\bf A}_{0}^{*}\bdot\pd{\bf x}{\varphi} \;=\; -\,\frac{e}{c}\,\psi  \;+\; p_{\|}\;b_{0\varphi} \;-\; J \left[ 2\,b_{0z} \;+\; \nabla\bdot\left(\frac{1}{2 B_{0}}\;\nabla\psi\right)\right],
\end{equation}
which contains higher-order guiding-center corrections \citep{Tronko_Brizard_2015}. In a careful numerical analysis of the exact particle orbits of energetic ions in a tokamak magnetic field, \cite{Belova_2003} have shown that the higher-order guiding-center corrections to the lowest-order guiding-center toroidal angular-momentum $P_{{\rm gc}\varphi}^{*} = -\,(e/c)\,\psi + p_{\|}\,b_{0\varphi} + \cdots$ play a crucial role in the guiding-center toroidal angular-momentum law (i.e., in the absence of electromagnetic-field perturbations). We note that it is a common practice to extract the dominant guiding-center contribution from $-\,(e/c)\psi$ by using the identity
\[ -\,\pd{}{t}\left( \int_{\bf P} {\cal J}_{\rm gy}\,F\;\frac{e}{c}\,\psi \right) - \nabla\bdot\left( \int_{\bf P} {\cal J}_{\rm gy}\,\dot{\bf X}\,F\;\frac{e}{c}\,\psi \right) \;=\; -\; \int_{\bf P} {\cal J}_{\rm gy}\,F\;\frac{e}{c}\,\dot{\psi} \;\equiv\; -\,\frac{1}{c}\;J_{\rm gy}^{\psi}, \]
where the radial velocity $\dot{\psi} \equiv \dot{\bf X}\bdot\nabla\psi$ is expressed in terms of the gyrocenter velocity $\dot{\bf X}$:
\begin{equation}
\dot{\psi} \;=\; \nabla\psi\bdot\left( {\bf E}_{\rm gy}^{*}\btimes\frac{c{\sf b}_{\rm gy}^{*}}{B_{{\rm gy}\|}^{**}} \;+\; \pd{K_{\rm gy}}{p_{\|}}\;\frac{{\bf B}_{\rm gy}^{*}}{B_{{\rm gy}\|}^{**}} \right).
\label{eq:psi_dot}
\end{equation}
Hence, we may now define $P_{{\rm gc}\varphi} \equiv P_{{\rm gc}\varphi}^{*} + (e/c)\,\psi$, and thus Eq.~\eqref{eq:Pphi*_final} becomes
\begin{equation}
\pd{{\cal P}_{{\rm gy}\varphi}}{t} + \nabla\bdot\left({\sf T}_{{\rm gy}}\bdot\pd{\bf x}{\varphi}\right) \;=\; \frac{1}{c}\;J_{\rm gy}^{\psi},
\label{eq:Pphi_final}
\end{equation}
where the toroidal angular-momentum density \eqref{eq:Pgy_can_phi} is now defined with $P_{{\rm gc}\varphi}$.

The second group in Eq.~\eqref{eq:Pgy_can_phi}, which appears because of the symplectic momentum perturbation $\vb{\Pi}_{\rm gy}$, contains the toroidal components of the perturbed $E\times B$ velocity and magnetic-flutter momentum, 
\begin{eqnarray}
\vb{\Pi}_{\rm gy}\bdot\pd{\bf x}{\varphi} &=& \epsilon\,\left(\langle{\bf E}_{1{\rm gc}}\rangle\btimes\frac{e\bhat_{0}}{\Omega_{0}} \;+\; \frac{p_{\|}}{B_{0}}\;\langle{\bf B}_{1\bot{\rm gc}}\rangle\right) \bdot\pd{\bf x}{\varphi} \nonumber \\
 &=& \frac{e}{B_{0}\Omega_{0}} \left( \langle{\bf E}_{1{\rm gc}}\rangle \;+\; \frac{p_{\|}\bhat_{0}}{mc}\btimes \langle{\bf B}_{1{\rm gc}}\rangle\right) \bdot\nabla\psi,
\end{eqnarray}
which can be expressed in terms of the radial component of the perturbed gyrocenter force, where we used the tokamak identity 
\begin{equation}
{\bf B}_{0}\btimes\partial{\bf x}/\partial\varphi \;=\; \nabla\psi. 
\label{eq:tokamak_id}
\end{equation}
The third group in Eq.~\eqref{eq:Pgy_can_phi} contains the toroidal component of the Minkowski electromagnetic momentum \citep{Abiteboul_2011}
\[ \frac{\mathbb{D}_{\rm gy}}{4\pi c}\btimes\epsilon\,{\bf B}_{1}\bdot\pd{\bf x}{\varphi} \;=\; \frac{1}{4\pi c}\,\left[\left(\epsilon\,{\bf E}_{1} \;+\frac{}{} 4\pi\,\mathbb{P}_{\rm gy}\right) \btimes \epsilon\,{\bf B}_{1}\right]\bdot\pd{\bf x}{\varphi}. \]
The partial time derivative of this term can be directly obtained from the toroidal component of Eq.~\eqref{eq:P_proof_2}.  We note that, in the electrostatic limit (i.e., in the absence of magnetic-field perturbations), we recover the flux-averaged gyrokinetic toroidal angular-momentum density previously derived (without guiding-center corrections, i.e., $P_{{\rm gc}\varphi} = p_{\|}\,b_{0\varphi}$) \citep{TSH_2007,Scott_Smirnov_2010, Brizard_Tronko_2011,Abiteboul_2011}.

Finally, we proceed with a flux-surface average \citep{Brizard_Tronko_2011}
\begin{equation}
\llbracket\;\cdots\;\rrbracket \;\equiv\; \frac{1}{{\cal V}}\;\oint\;(\cdots)\;{\cal J}_{\psi}\;d\vartheta\,d\varphi,
\label{eq:flux_av}
\end{equation}
where ${\cal V}(\psi) \equiv \oint\;{\cal J}_{\psi}\;d\vartheta\,d\varphi$ is the surface integral of the magnetic-coordinate Jacobian ${\cal J}_{\psi} \equiv (\nabla\psi\btimes\nabla\theta\bdot\nabla\varphi)^{-1} = 1/B_{0}^{\theta}$. The flux-surface average \eqref{eq:flux_av} satisfies the property
\begin{equation}
\llbracket\nabla\bdot{\bf C}\rrbracket \;\equiv\; \frac{1}{{\cal V}}\;\pd{}{\psi}\left( {\cal V}\;\left\llbracket\frac{}{} {\bf C}\bdot\nabla\psi\right\rrbracket
\right)
\label{eq:mag_surf_def}
\end{equation}
for any vector field ${\bf C}$. In a time-independent axisymmetric tokamak geometry, we note that $\partial/\partial t$ also commutes with magnetic-surface averaging.  The magnetic surface-averaged gyrokinetic canonical angular-momentum conservation law \eqref{eq:Pphi_final} becomes
\begin{equation}
\pd{\llbracket{\cal P}_{{\rm gy}\varphi}\rrbracket}{t} \;+\; \frac{1}{{\cal V}}\;\pd{}{\psi}\left( {\cal V}\;\left\llbracket\frac{}{} T^{\psi}_{{\rm gy}\varphi}\right\rrbracket\right) \;=\; \frac{1}{c}\,\llbracket J_{\rm gy}^{\psi}\rrbracket,
\end{equation}
where $T^{\psi}_{{\rm gy}\varphi} \equiv \nabla\psi\bdot{\sf T}_{\rm gy}^{*}\bdot\partial{\bf x}/\partial\varphi$ is defined as
\begin{equation}
T^{\psi}_{{\rm gy}\varphi} \;=\;  \int_{\bf P} {\cal J}_{\rm gy}\,F\;\dot{\psi}\,\left( P_{{\rm gc}\varphi} \;+\; \vb{\Pi}_{\rm gy}\bdot\pd{\bf x}{\varphi} \right) \;-\; \frac{\epsilon}{4\pi} \nabla\psi\bdot \left( \mathbb{D}_{\rm gy}\,{\bf E}_{1} \;+\frac{}{} {\bf B}_{1}\,\mathbb{H}_{\rm gy}\right) \bdot\pd{\bf x}{\varphi},
\label{eq:Tgy_can_psi}
\end{equation}
where we have used $\nabla\psi\bdot\partial{\bf x}/\partial\varphi = 0$ and $\dot{\psi}$ is given in Eq.~\eqref{eq:psi_dot}. We note that, using the tokamak identity \eqref{eq:tokamak_id}, the third term in Eq.~\eqref{eq:Tgy_can_psi}, which contains the polarization term derived by \cite{McDevitt_2009} in the parallel limit, can be expressed as
\[  \nabla\psi\bdot\left( \frac{\epsilon}{4\pi}\,\mathbb{D}_{\rm gy}\,{\bf E}_{1} \right)\bdot\pd{\bf x}{\varphi} \;=\; \pd{\bf x}{\varphi}\bdot\left[ \frac{\epsilon}{4\pi}\,\left(\mathbb{D}_{\rm gy}\btimes{\bf B}_{0}\right)\,{\bf E}_{1} \right]\bdot\pd{\bf x}{\varphi}, \]
and similarly for the fourth term. Similar terms have appeared in the toroidal angular-momentum transport analysis of \cite{Parra_Catto_2010}.

\section{\label{sec:Sum}Summary}

The energy-momentum and toroidal angular-momentum conservation laws of two gauge-free gyrokinetic Vlasov-Maxwell models were derived by Noether method under the assumption of a time-independent and axisymmetric equilibrium magnetic field.  The explicit proof of these conservation laws highlights the roles played by the equilibrium magnetized plasma and the electromagnetic-field fluctuations that perturb it. In addition, we also demonstrated how the gyrokinetic Noether momentum equation 
\eqref{eq:momentum} follows exactly from the gyrokinetic Vlasov-Maxwell equations. Hence, a gyrokinetic parallel-momentum transport equation can be derived explicitly without proceeding through a gyrokinetic Vlasov-moment approach. 

The proofs presented in Sec.~\ref{sec:symp_laws} also show how gyrokinetic models can be simplified without jeopardizing the energy-momentum conservation laws. For example, \cite{EH_2020} considered the simplified gyrocenter kinetic energy \eqref{eq:Kgy_BB} for the Hamiltonian gyrokinetic model \citep{Burby_Brizard_2019} obtained by omitting the guiding-center electric-dipole moment $\vb{\pi}_{\rm gc}$. This omission yields simplified expressions for  the gyrocenter polarization and magnetization \eqref{eq:Pol_gy}-\eqref{eq:Mag_gy}, without sacrificing energy and angular-momentum conservation. 

In the gauge-free symplectic gyrokinetic model considered by \cite{Brizard_2020}, it is possible to truncate the gyrocenter kinetic energy \eqref{eq:Kgy_B} at first order in $\epsilon$, thereby eliminating the corrections $\vb{\pi}_{2}$ and $\mu\,{\bf B}_{1}/B_{0}$ in the gyrocenter polarization and magnetization \eqref{eq:Pol_gy}-\eqref{eq:Mag_gy}, which arise from functional derivatives of the second-order gyrocenter Hamiltonian.

Finally, we note that an exact toroidal angular-momentum conservation for the gyrokinetic Vlasov-Maxwell equations is obtained even though the gyrokinetic stress tensor is manifestly asymmetric. In contrast to the guiding-center Vlasov-Maxwell equations, where the interplay between ponderomotive, polarization, and magnetization effects results in a symmetric guiding-center stress tensor (as required because the magnetic field is not split into background and perturbed components), the case of the standard gyrokinetic splitting of the magnetic field into background and perturbed components requires an asymmetric gyrokinetic stress tensor, as can be seen from Eq.~\eqref{eq:Pphi}. The recent work by \cite{Chen_2020} and \cite{Sugama_2021} may pave the way to a nonlinear gyrokinetic theory with full electromagnetic effects without field splitting, from which a symmetric stress tensor will arise (but only a careful analysis of ponderomotive, polarization, and magnetization effects is carried out).

\acknowledgments

Part of the work presented here was carried out as part of a collaboration with the {\sf ELMFIRE} numerical simulation group at Aalto University (Finland). The Author acknowledges support from the National Science Foundation under contract No.~PHY-1805164. In addition, the Author reports no conflict of interest.

 \bibliographystyle{jpp}

\bibliography{symp}

\end{document}